\def\astroph{1}
  \ifnum\astroph=0
     \documentclass[12pt,preprint]{aastex}
  \else
    \documentclass{emulateapj}
   \fi

  \usepackage{epsfig,amsmath,natbib}
  \usepackage[dvips]{color}
  \usepackage{verbatim}

\begin{document}
 \slugcomment{Submitted to ApJ}
\shorttitle{LY$\alpha$ RT IN COSMOLOGICAL SIMULATIONS USING AMR}
\shortauthors{LAURSEN ET AL.}
\title{Lyman $\alpha$ Radiative Transfer in Cosmological Simulations\\
       using Adaptive Mesh Refinement}
\author{Peter Laursen\altaffilmark{1},
        Alexei O.~Razoumov\altaffilmark{2}
        and Jesper Sommer-Larsen\altaffilmark{3,1}}
%
%
\altaffiltext{1}{Dark Cosmology Centre, Niels Bohr Institute, University of
                 Copenhagen, Juliane Maries Vej~30, DK-2100, Copenhagen {\O},
                 Denmark; email: pela@dark-cosmology.dk}
\altaffiltext{2}{Institute for Computational Astrophysics, Dept.~of Astronomy
                 \& Physics, Saint Mary's University, Halifax, NS, B3H3C3,
                 Canada; email: razoumov@ap.smu.ca}
\altaffiltext{3}{Excellence Cluster Universe, Technische Universit\"at
                 M\"unchen, Boltzmannstra\ss e 2, D-85748 Garching,
                 Germany; email:\\ jslarsen@astro.ku.dk}

\begin{abstract}
A numerical code for solving various Ly$\alpha$ radiative transfer (RT)
problems is presented.
The code is suitable for an arbitrary, three-dimensional distribution of
Ly$\alpha$ emissivity, gas temperature, density, and velocity field.
Capable of handling Ly$\alpha$ RT in an adaptively refined grid-based
structure, it enables detailed investigation of the effects of
clumpiness of the interstellar (or intergalactic) medium.
The code is tested against various geometrically and physically idealized
configurations for which analytical solutions exist, and subsequently
applied to three different simulated high-resolution ``Lyman-break galaxies'',
extracted from high-resolution cosmological simulations at redshift $z = 3.6$.
Proper treatment of the Ly$\alpha$ scattering reveals a diversity of surface
brightness (SB) and line profiles. Specifically, for a given galaxy
the maximum observed SB can vary by an order of magnitude,
and the total flux by a factor of 3--6, depending on the viewing angle.
This  may provide an explanation for differences in observed properties of
high-redshift galaxies, and in particular a possible physical link
between Lyman-break galaxies and regular Ly$\alpha$ emitters.
\end{abstract}

\section{Introduction}
The significance of the Ly$\alpha$ emission line as a probe of the
high-redshift Universe has long been established.
In their classic paper, \citet{par67} suggested how detection of young galaxies
would be feasible using the Ly$\alpha$ line. Nevertheless, for almost three
decades only a few Ly$\alpha$ emitters (LAEs) were discovered
\citep[see, e.g.,][]{djo92}. Several theories were proposed to explain the
many null results, e.g.~suppression of the Ly$\alpha$ line due to metals
\citep{mei81}, absorption by dust \citep{har88} and lower-than-expected
formation of massive stars \citep{val93}.

However, as surveys eventually were able to go deeper, and as searching wide
regions on the sky became feasible, large numbers of high-redshift,
star-forming galaxies were discovered. Notable surveys include
the Large Area Lyman Alpha survey \citep[e.g.,][]{rho00} and
the Subaru Deep Field survey \citep[e.g.,][]{tan05}.
Currently, redshifts of LAEs up to $z \simeq 7$
\citep{iye07} have been reached, while galaxies detected via $z$-band dropout
observations have reached $z \sim 8$ \citep{bou04},
but in time deeper observations will be realized, with the advent of, e.g., the
Ultra-VISTA project
\citep[to be launched ultimo this year, reaching $z=8.8$; ][]{dun07}
and the James Webb Space Telescope
\citep[to be launched in 2013; e.g.,][]{gar06}.

Besides contributing to our understanding of the overall
structure and evolution of the Universe, much insight into the properties and
formation of the galaxies themselves, the fundamental building blocks of our
Universe, has now been gained from these high-redshift surveys.
Thus, numerous characteristics of the LAEs, such as their
density \citep{hu98}, clustering properties \citep{ouc03}, and luminosity
function \citep{hu04}, have been subject to investigation.

Radiative transfer (RT) is playing an increasingly important role in numerical
astrophysics and cosmology.
This is particularly true in the case of Ly$\alpha$. Due to the resonant
scattering nature of Ly$\alpha$ radiation, and due to the fact that neutral
hydrogen is abundantly present in the interstellar medium (ISM) and the
intergalactic medium (IGM),
the history of a Ly$\alpha$ photon, from the moment it is created in a
high-redshift galaxy to the time it is captured by a telescope, is not trivial.
In general, analytical solutions of RT problems
are obtainable only in very idealized cases.
By far, most of the work done on the subject has been concerned with the
emerging spectrum from an isothermal,
homogeneous medium of plane-parallel or spherical symmetry
\citep[e.g.][]{aue65,ave68,pan73,ahn01,ahn02,zhe02}. Some allow for isotropic
velocities \citep[e.g.][]{car72,nat86,loe99,dij06}, and some include simple
models for dust \citep[e.g.][]{bon79,ahn00,han06,ver06,ver07}.
However, even though the results of this work
have improved tremendously our knowledge of many physical processes, they do
not capture the complexity and diversity of realistic, astrophysical situations
where velocities can be quite chaotic, and densities and temperatures can vary
by many orders of magnitude over relatively small distances.

To this end, a few codes with varying aims have been constructed, capable of
performing realistic RT for arbitrary
distributions of source Ly$\alpha$ emission, neutral hydrogen density,
temperature and velocity resulting from cosmological
simulations so as to yield the spectrum and the spatial distribution of the
escaping photons \citep{can05,tas06a,kol06}. Also
\citet{ver06} have presented a similar code, although it has not been applied
to cosmological simulations.

Although the work carried out in this paper is largely inspired by previous
studies, it improves on earlier works in several different ways:
most importantly, as in the case of the code of \citet{tas06a}, our code is
capable of working with physical data on an adaptively refined mesh, as
opposed to a regular grid. 
Since gas clumping affects the photon escape probability, very high
resolution is desired. Our study will be restricted to
quite evolved galaxies, on the kpc scale.
Furthermore, in addition to studying the emergent spectrum and surface
brightness distribution, we will investigate the effect of viewing the systems
from different angles.

The rest of the paper is organized as follows: in \S\ref{sec:rt}, a basic
theory of Ly$\alpha$ RT is briefly described.
The equations presented in this section
will serve as a basis for understanding and testing the code.
A detailed description of the principles of the RT code is given in
\S\ref{sec:code}, and in \S\ref{sec:tests} tests of the code against various
analytical solutions are presented.
A semianalytical acceleration scheme is derived in \S\ref{sec:acc}, and
in \S\ref{sec:app} the code is applied to three different galaxies extracted
from high-resolution cosmological simulations. Finally, a discussion of the
results is given in \S\ref{sec:disc}.

\section{Resonant Scattering Radiative Transfer}
\label{sec:rt}

The first attempts to predict the diffusion of Ly$\alpha$ were made under the
assumption of coherent scattering in the observers frame \citep{amb32,cha35}.
The probability of interaction between a photon and an atom at rest with
respect to the reference frame in which the frequency $\nu$ of the photon is
measured is described by the line profile $\phi(\nu)$;
this was known to be given
by the sharply peaked natural (Lorentzian) line profile $\mathcal{L}$ of
width $\Delta\nu_{\mathrm{L}} = 9.936\times10^{7}$ Hz around the line center
frequency $\nu_0 = 2.466\times10^{15}$ Hz.

Several physical quantities that may or may not be directly observable have
been the subject of interest, e.g.~the average number
$\langle N_{\mathrm{scat}} \rangle$ of
scatterings required to escape the medium (to determine the probability of a
photon being destroyed by dust grains, or by collisions of the scattering atom
with other atoms while being in the excited state)
and the shape of the emergent spectrum.
Due to the complexity of the problem, the physical configurations
investigated have traditionally been restricted to problems in which
photons are emitted in the center of a homogeneous, isothermal cloud
which is either spherically symmetric, or infinite in two directions
and has a finite extension in the third direction (a plane-parallel ``slab'').
Denoting by $\tau_0$ the optical depth for a photon
in the line center from the
initial point of emission to the edge of the gaseous cloud, from pure random
walk considerations one would naively infer
$ \langle N_{\mathrm{scat}} \rangle \sim \tau_0^2$.  Accordingly, the medium
would not have to be very opaque in order for the destruction processes of
Ly$\alpha$ to become significant. 
\ \\
\ \\

\subsection{Thermal Broadening of the Line}
\label{sec:therm}

\citet{hen40} and \citet{spi44} acknowledged the fact that scattered photons
undergo a change in frequency due to thermal Doppler broadening of the
scattering atoms.
In the following, to simplify notation the frequency of the photon is
parametrized through $x = (\nu - \nu_0)/\Delta\nu_{\mathrm{D}}$,
where $\Delta\nu_{\mathrm{D}} = (v_{\mathrm{th}}/c)\nu_0$ is the width of the
Doppler (Gaussian) profile, with
$v_{\mathrm{th}} = (2 k_B T / m_{\mathrm{H}})^{1/2}$ being the thermal
atom velocity dispersion (times $\sqrt{2}$) and the rest of the variables
having their usual meaning.
In terms of these quantities, with $\phi(\nu)d\nu = \phi(x)dx$ the normalized
thermal line profile is
\begin{equation}
\label{eq:Gx}
\mathcal{G}(x) = \frac{1}{\sqrt{\pi}} e^{-x^2},
\end{equation}
while the natural line profile is
\begin{equation}
\label{eq:Lx}
\mathcal{L}(x) = \frac{a}{\pi} \frac{1}{x^2 + a^2},
\end{equation}
where $a \equiv \Delta\nu_L/2\Delta\nu_D$ is the relative line width.
The resulting (Voigt) profile is a convolution between the two, but due to the
smallness of $a$, the center of the profile is entirely dominated by
$\mathcal{G}$.

Relying on these considerations,
\citet{zan49,zan51} argued that, in each scattering, the frequency of the
Ly$\alpha$
photon would undergo \emph{complete redistribution} over the Doppler line
profile, i.e.~there is no correlation between the frequency $x_i$ of the
incoming and $x_f$ of the outgoing photon,
and the probability that $x < x_f < x+dx$ is $\phi(x_f)dx$. In this picture,
the photon still executes a random walk, but at each scattering there is a small
possibility that it will be redistributed so far into the wing as to render the
medium optically thin and thus allow escape.
This reduces $\langle N_{\mathrm{scat}} \rangle$ significantly, and the result
was later
verified numerically for intermediate optical depths ($\tau_0 \sim 10^4$) by
\citet{koe56}.

Still based on the assumption of isotropic scattering,
\citet{unn52a,unn52b} calculated an ``exact redistribution'' formula
$q(x_i,x_f)$, giving the probability distribution of $x_f$ as a function of
$x_i$. With this result,
\citet{ost62} found that in the wings, the rms frequency shift
$(\Delta x)_{\mathrm{rms}}$ per scattering is
\begin{equation}
\label{eq:rms}
(\Delta x)_{\mathrm{rms}} = 1,
\end{equation}
and the mean shift $\langle \Delta x \rangle$ per scattering is
\begin{equation}
\label{eq:meanx}
\langle \Delta x \rangle = -1/|x|,
\end{equation}
i.e.~there is a tendency to drift toward the line center.
Thus, a photon at frequency
$x \gg 1$ will execute a nearly random walk in frequency, returning to the
core in $N_{\mathrm{scat,ret.}} \sim x^2$ scatterings.

From $q(x_i,x_f)$, \citet{ost62} found that
$\langle N_{\mathrm{scat}} \rangle \propto \tau_0$
for moderate optical depths. However, he
argued that for some limiting large optical depth --- which he was not able to
calculate due to the lack of ``sufficiently large digital computers'' ---
$x_f$ is
so large that the photon will execute a random walk also in real space, whence
in this case $\langle N_{\mathrm{scat}} \rangle \propto \tau_0^2$.

Nonetheless, applying the method of \citet{fea64}, \citet{ada72} found
numerically that also for extremely large
optical depths ($\tau_0$ up to $10^8$),
$\langle N_{\mathrm{scat}} \rangle \propto \tau_0$.
Although he could not
prove it rigorously, he was able to give a heuristic argument on physical
grounds for this behavior.

\subsection{Neufeld Solution}
\label{sec:neufeld}

The result was proven the subsequent year by
\citet{har73}:
inspired by \citet{unn55}, utilizing the Eddington approximation --- which
implies that the radiation field is everywhere nearly isotropic, but with a
small net outward flow --- and
expanding the redistribution function as formulated by \citet{hum62} to
second order, \citet{har73} obtained a diffusion equation for the angular
averaged intensity $J(\tau,x)$ within a (nonabsorbing) slab of extremely
large optical depths (defined\footnote{Note that in \citeauthor{har73}'s papers, as well
as most coeval authors', the optical depth at frequency $x$ is defined as
$\tau_x=\tau_0 \phi(x)$, whereas in our definition $\tau_x=\tau_0 H(a,x)$.
Since $H(a,x)=\sqrt{\pi}\phi(x)$, this implies that
$\tau_{\mathrm{Harrington}} = \sqrt{\pi}\tau_{\mathrm{us}}$. This definition
has been chosen to follow more recent studies.} as
$a\tau_0 \ge 10^3 / \sqrt{\pi}$, or $\tau_0 \ge 1.2\times10^6$ for
$T = 10^4$ K).

With the photons emitted isotropically from a central source
emitting 1 photon per unit time, i.e.~$1/4\pi$ photons per unit time per
steradian, an initial frequency $x_{\mathrm{inj}} = 0$, and scatterings assumed
to be dominated by isotropic wing scatterings, \citet{har73}
obtained an expression for the emergent spectrum. \citet{neu90} gave a more
general solution to the problem, allowing for the destruction of photons and
the injection at any initial optical depth $\tau$ in the slab, with arbitrary
initial frequency.
For centrally\footnote{\citeauthor{neu90} assumed that the photons
are emitted from a thin layer inside the slab, parallel to the surface.
However, for reasons of symmetry, we may as well assume that they are emitted
from a single point.} (at $\tau = 0$) emitted radiation in a nonabsorbing
medium, the solution at the surface, i.e.~at $\tau = \pm\tau_0$, is
\begin{equation}
\label{eq:neufeld}
J(\pm\tau_0,x) = \frac{\sqrt{6}}{24} \frac{x^2}{\sqrt{\pi}a\tau_0}
 \frac{1}{\cosh\big[\sqrt{\pi^3/54}\, (x^3 - x_{\mathrm{inj}}^3)/a\tau_0 \big]}.
\end{equation}
With perhaps some injustice, we will refer to Eq.~(\ref{eq:neufeld}) as the
``Neufeld solution'', even when $x_{\mathrm{inj}} = 0$, in which case it
reduces to the
result of \citet{har73}. The profile is normalized to $1/4\pi$ and exhibits
two bumps, symmetrically centered on $x=0$ and drifting further
apart for increasing $a\tau_0$. Note that it solely
depends on the product $a\tau_0$, and that the physical size of the gaseous
system does not enter the equation. A higher density is compensated for
by a higher temperature, since
$a \tau_0 \propto
(\Delta\nu_D^{-1})
(n_{\textrm{{\scriptsize H}{\tiny \hspace{.1mm}I}}} \Delta\nu_D^{-1}) \propto
n_{\textrm{{\scriptsize H}{\tiny \hspace{.1mm}I}}}/T$ (at a given size).
The physical explanation for
this is that the denser the medium is, the further into the wing the
photons have to drift. Meanwhile, a higher temperature --- and a resulting
higher velocity dispersion of the atoms --- will make the medium less
opaque to radiation, since this means fewer atoms with a velocity matching the
frequency of the photons.

Setting $\partial J/\partial x = 0$ , \citet{har73} showed that the emergent
spectrum has its maximum at
\begin{equation}
\label{eq:xm}
x_m = \pm 1.066 (a\tau_0)^{1/3},
\end{equation}
while the average number of scatterings that
a photon undergoes before escaping the slab was shown to be
\begin{equation}
\label{eq:Nscat}
\langle N_{\mathrm{scat}} \rangle = 1.612 \tau_0.
\end{equation}

Except for numerical factors of order unity, \citet{dij06} derived similar
expressions for the emergent spectrum and the number of scatterings for photons
escaping a static, isothermal, homogeneous sphere of gas.
Furthermore, the spectrum for an
isotropically expanding (as in Hubble flow) or contracting (as in a
gravitational collapse) medium, but with no thermal motion, was examined
analytically by \citet{loe99}.
Evidently, all of the configurations considered so far are highly idealized
compared to realistic, astrophysical situations and for more general
geometries and velocities, analytic solutions are not obtainable.
Nevertheless, they provide valuable and
at least qualitative insight into the characteristics of young galaxies,
the ISM and IGM,
H\,\textsc{i} envelopes surrounding hot stars, etc.
Moreover, they offer direct means of testing numerical methods (see
\S\ref{sec:tests}).

\section{The Code}
\label{sec:code}

The transfer of the Ly$\alpha$ photons is conducted using the 3D adaptive mesh
refinement (AMR) Monte Carlo code {\sc MoCaLaTA}.
Except for a few numerical improvements, in particular the acceleration scheme
described in \S\ref{sec:acc}, the code resembles the one presented
in \citet{lau07}, with one significant advance: it is now capable of assuming
an adaptively refined mesh, to an arbitrary level of refinement. This allows
for the opportunity to study the effect of the clumpiness of the
ISM on the radiative transfer in great detail.

The principles of the code were briefly explained in \citet{lau07}. In the
following we give a more elaborate description of how the RT is realized.

\subsection{Ly$\alpha$ Emission}
\label{sec:em}

The physical volume of interest is discretized on a base grid, typically
consisting of $128^3$ cells. Cells may be subdivided into eight subcells which,
in turn, may be
further refined. The refinement criterion is usually taken to be density, but
can in principle be any condition, e.g.~density gradient, velocity, etc.
If the underlying cosmological simulation is particle based,
as is the case in the present study, the physical parameters of interest
are first interpolated onto the grid.
 
Each cell contains information about the Ly$\alpha$ luminosity
$L_{\mathrm{Ly}\alpha}$ and the density
$n_{\textrm{{\scriptsize H}{\tiny \hspace{.1mm}I}}}$ of neutral hydrogen, as
well as the temperature $T$ and the three-dimensional peculiar velocity field
$\mathbf{v}_{\mathrm{bulk}}$ of the gas elements.
The ratio of $L_{\mathrm{Ly}\alpha}$ of a given cell to the total luminosity
$L_{\mathrm{tot}}$ of all cells determines the probability of a photon being
emitted from that particular cell. The initial position $\mathbf{x}_i$ of the
photon is a random location in the cell. In the reference frame of the emitting
atom, the photon is injected with a frequency $x_{\mathrm{nat}}$, given by the
distribution $\mathcal{L}(x)$ (Eq.~\ref{eq:Lx}).
The atom, in turn, has a velocity $\mathbf{v}_{\mathrm{atom}}$ in the reference
frame of the gas element
drawn from a thermal profile of Doppler width $\Delta\nu_{\mathrm{D}}$.
Measuring atom velocities in terms of Doppler widths,
$\mathbf{u} = \mathbf{v}_{\mathrm{atom}}/v_{\mathrm{th}}$, each component $u_i$
is then distributed according to $\mathcal{G}(u_i)$, given by
Eq.~(\ref{eq:Gx}).

The initial direction $\mathbf{\hat{n}}_i$ of the photon follows an isotropic
probability distribution.
To first order in $v/c$, this is true in all relevant reference frames,
and a Lorentz transformation to the reference frame of
the gas element then yields the initial
frequency $x_i = x_{\mathrm{nat}} + \mathbf{u} \cdot \mathbf{\hat{n}}_i$.

For photons emitted in the dense, star-forming regions, it makes no difference
whether $x_i$ is calculated in the above manner or simply set
equal to zero. However, when studying large volumes of space, a nonvanishing
fraction of the Ly$\alpha$ photons may be produced through cooling radiation,
which
also takes place well away from the star-forming regions of the galaxy.
In these environments, whereas the probability of a photon with $x = 0$
escaping is still
extremely small, being injected one or two Doppler widths away from line center
may allow the photon to escape.

\subsection{Propagation of the Radiation}
\label{sec:prop}

The optical depth $\tau$ covered by the photon before it is scattered is
governed by the probability density function $P(\tau) = e^{-\tau}$, and after
initial emission and all subsequent scatterings, a random value of $\tau$ is drawn
from $P(\tau)$.
This optical depth is converted into a physical distance
$r = \tau / n_{\textrm{{\scriptsize H}{\tiny I}}} \sigma_x$. In the reference
frame of the gas, the cross-section $\sigma_x$ of the atom responsible for the
scattering event is given by the Voigt profile, resulting in
\begin{equation}
\label{eq:xsec}
\sigma_x = f_{12} \frac{\sqrt{\pi} e^2}{m_e c \Delta\nu_{\mathrm{D}}} H(a,x),
\end{equation}
where $f_{12} = 0.4162$ is the Ly$\alpha$ oscillator strength, $m_e$ is the
mass of the electron, and
\begin{equation}
\label{eq:H}
H(a,x) = \frac{a}{\pi} \int_{-\infty}^{+\infty}
         \frac{e^{-y^2}}{(x-y)^2 + a^2} dy
\end{equation}
is the Voigt function.
This function can be approximated by a Gaussian in the core and a power law
in the wing. However, in the transition domain between core and wing,
either approximation is poor. Hence, instead we use the
analytical fit \citep{tas06a}
\begin{equation}
\label{eq:Haxtas}
H(a,x) = q\sqrt{\pi} + e^{-x^2},
\end{equation}
where
\begin{equation}
\label{eq:q}
q = \left\{ \begin{array}{ll}
0                                     & \textrm{for } \zeta \le 0\\
\left(1 + \frac{21}{x^2}\right) 
          \frac{a}{\pi(x^2 + 1)} \Pi(\zeta) & \textrm{for } \zeta > 0,
\end{array} 
\right.
\end{equation}
with
%
$\zeta = (x^2 - 0.855) / (x^2 + 3.42)$ and
$\Pi(\zeta) = 5.674\zeta^4 - 9.207\zeta^3 + 4.421\zeta^2 + 0.1117\zeta$.
This is an excellent approximation for all frequencies at temperatures above
2 K.

All physical parameters entering the equations above are of course given by
the cell in which the photon is presently located; the \emph{host cell}. The new
position is then $\mathbf{x}_f = \mathbf{x}_i + r\hat{\mathbf{n}}_i$. However,
since in general the physical conditions vary from cell to cell,
if $\mathbf{x}_f$ is outside the host cell, the photon is placed at the point
$\mathbf{x}_{\mathrm{cut}}$ of intersection with the face of the
cell and the above calculation is redone with the parameters of the new cell.
Part of the originally assigned value of $\tau$ has already been ``spent'', so
the remaining optical depth to be traveled is now
\begin{equation}
\label{eq:tau2tau}
\tau = \tau_{\mathrm{orig.}}
     - \big|\mathbf{x}_{\mathrm{cut}} - \mathbf{x}_i\big|
     \big( n_{\textrm{{\scriptsize H}{\tiny I}}} \sigma_x \big)_{\mathrm{prev.cell}}.
\end{equation}
%
The frequency of the photon is Lorentz transformed to the
bulk velocity of the new cell.

In contrast to a  regular grid, in an AMR grid a given cell will not in general
have a unique neighbor.
The cells are structured in a nested grid, where a refined cell is the
``parent'' of eight ``child'' cells which, in turn, may or may not be refined.
The new host cell of the photon is then determined by walking
up and down the hierarchical tree structure.

\subsection{Scattering}
\label{sec:scat}

\begin{figure}
\epsscale{0.8}
\plotone{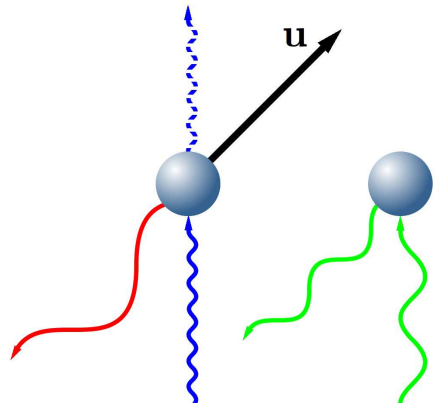}
\caption{Illustration of the mechanism responsible for the frequency
         shift of a scattered photon. In the reference frame of an
         external observer (\emph{left}), a photon blueward of the line center
         (\emph{blue solid}) is scattered by an atom receding in such manner that the
         component $u_{||}$ of its velocity $\mathbf{u}$ along the direction of
         the photon matches closely the frequency $x$. In the reference
         frame of the atom (\emph{right}), the photon then seems close to the
         line center (\emph{green}). Except for a minute recoil effect, the
         photon leaves the atom with the same frequency. However, to the
         external observer, if the photon is scattered in a direction
         opposite the atom's motion (red), it will be redshifted. Only
         if by chance it is scattered in exactly the original direction
         (\emph{dotted blue}), its frequency remains unaltered. For $|x| \gg 1$, the
         number of atoms with $u_{||} \simeq x$ is so small that the photon is
         most likely to be scattered by a low-velocity atom. In this case, no
         matter in which direction the photon is scattered the motion of the
         atom will not shift the frequency significantly.}
\label{fig:RefFrame}
\end{figure}

When the initially assigned $\tau$ is ``used up'', the photon is scattered.
It must be emphasized that the discussed broadening of the line and the
corresponding uncertainty in energy does \emph{not} imply that a photon of a
given energy can be absorbed, and subsequently re-emitted with a different
energy. Indeed, this would be possible had the energy of the ground state been
associated with an uncertainty in energy as well. However, since the lifetime
of this state is effectively infinite, its energy is well-defined.
Except for a small recoil effect, the scattering is coherent in the reference
frame of the atom. However, to an external observer the nonzero velocity of
the scattering atom will, in general, add a frequency shift to the photon.
Figure \ref{fig:RefFrame} shows a qualitative interpretation of how the
Doppler shift arises.
Since the frequency determines the opacity of the gas, the exact value of the
velocity is important.
In the directions perpendicular to $\mathbf{\hat{n}}_i$,
the velocities $u_{\perp1,2}$ will follow a Gaussian
distribution $\mathcal{G}(u_{\perp1,2})$.
However, due to the resonance nature of the scattering event,
the velocity $u_{||}$ parallel to $\mathbf{\hat{n}}_i$  depends on $x$. Thus,
the probability distribution $\mathcal{G}(u_{||})$ must be convolved with the
probability $\mathcal{L}(x-u_{||})$ of the atom being able to scatter the
photon.
For small values of $|x|$, being scattered by an atom with $u_{||} = x$ is
highly favored. For large values of $|x|$ the abundance of these atoms
decrease as $e^{-x^2}$, so that scattering by ``slow'' atoms becomes more
probable, even though in reference frame of these atoms the photon is far in
the wing. The resulting normalized probability distribution is
\begin{equation}
\label{eq:fu}
f(u_{||}) = \frac{a}{\pi H(a,x)} \frac{e^{-u_{||}^2}} {(x-u_{||})^2 + a^2}.
\end{equation}
Since Eq.~(\ref{eq:fu}) is not analytically integrable,
$u_{||}$ is generated from this distribution by means of the \emph{rejection
method} \citep{pre92}:
a random value of a comparison function that \emph{is} integrable and lies
everywhere above $f(u_{||})$ is found, and accepted if a second random
number\footnote{Random numbers in the interval $[0,1]$ are generated by means
of the subroutine {\tt ran1} \citep{pre92}.} $\mathcal{R} \in [0,1]$ (a
``univariate'') is
less that the ratio of the two functions. Due to the peculiar shape of
$f(u_{||})$ (see Fig.~\ref{fig:uII}),
following \citet{zhe02}, \emph{two} comparison functions are used.
For the
wide range of temperatures and frequencies involved we find that a
satisfactory average acceptance-to-rejection ratio of order unity is
achieved for
\begin{equation}
\label{eq:u0}
u_0 = \left\{ \begin{array}{ll}
0                           & \textrm{for }
                              0 \phantom{.2}\le x < 0.2\\
x - 0.01 a^{1/6} e^{1.2x}   & \textrm{for }
                              0.2 \le x < x_{\mathrm{cw}}(a)\\
4.5                         & \textrm{for }
                              \phantom{0.2| \le } x \ge x_{\mathrm{cw}}(a)
\end{array}
\right.
\end{equation}
as the value $u_0$ separating the two comparison functions.
Here $x_{\mathrm{cw}}$ defines the boundary between the core and the wings of
the Voigt profile, i.e.~where $e^{-x^2}/\sqrt{\pi} = a/\pi x^2$.
The solution to this equation can be approximated as
\begin{equation}
\label{eq:xcw}
x_{\mathrm{cw}}(a) = 1.59 - 0.60 \log a - 0.03 \log^2 a.
\end{equation}

When $x \simeq 0$, the photon barely diffuses spatially. Only when
it has diffused sufficiently far in frequency space will it be able to make
a large journey in real space.
The photon may scatter thousands or even hundreds of thousands of times before
entering the wing of the line profile. These
scatterings are insignificant in the sense that they do not contribute to any
important displacement in neither space nor frequency. Hence, we may as well
skip them altogether and go directly to the first scattering that pushes the
photon into the wing.
This highly efficient acceleration of the code is achieved following
\citet{ahn02}: if $|x|$ is less than some critical value $x_{\mathrm{crit}}$,
$u_{\perp1,2}$ is drawn from a truncated Gaussian so as
to favor fast moving atoms and artificially push the photon back in the wing.
The resulting random velocity generator can be written \citep{dij06} as
\begin{equation}
\label{eq:u12dij}
\left. \begin{array}{lll}
u_{\perp,1} & = & \left( x_{\mathrm{crit}}^2 - \ln \mathcal{R}_1 \right)^{1/2}
                  \cos2\pi\mathcal{R}_2\\
u_{\perp,2} & = & \left( x_{\mathrm{crit}}^2 - \ln \mathcal{R}_1 \right)^{1/2}
                  \sin2\pi\mathcal{R}_2,\\
\end{array}
\right.
\end{equation}
where $\mathcal{R}_1$ and $\mathcal{R}_2$ are two univariates.

However,
the value $x_{\mathrm{crit}}$ is not simply equal to $x_{\mathrm{cw}}$, since
for a nondense medium, a core scattering can in fact be associated with a
considerable spatial journey, while for clouds of extremely high density
even scatterings
in the inner part of the wing may be neglected. Moreover, the exact value of
$x_{\mathrm{crit}}$ is actually quite important; this acceleration scheme can
decrease the
computational execution time by \emph{several orders of magnitude} but
too high values will push the photons unnaturally far out in the wings,
leading to incorrect results.
From the Neufeld solution we know that the important
parameter is the product $a\tau_0$. Correspondingly, we expect
$x_{\mathrm{crit}}$ to be a function of the value of $a\tau_0$ in the current
cell. Indeed, it is found that the value
\begin{equation}
\label{eq:xcrit}
x_{\mathrm{crit}} = \left\{ \begin{array}{ll}
0                               & \textrm{for } a\tau_0 \le 1\\
0.02 e^{\xi \ln^\chi\! a\tau_0} & \textrm{for } a\tau_0 > 1,
\end{array}
\right.
\end{equation}
where $(\xi,\chi) = (0.6,1.2)$ or $(1.4,0.6)$ for $a\tau_0 \le 60$ or
$a\tau_0 > 60$,
respectively, can be used without affecting the emergent spectrum in both
various tests (\S\ref{sec:tests}) and realistic situations (\S\ref{sec:app}).
Of course, if the photon is already in the wing, the proper Gaussian velocity
distribution is used, i.e. $x_{\mathrm{crit}} = 0$.

The final frequency $x_f$ of the scattered photon (in the reference frame of the
fluid element) depends on direction in which the photon is scattered, given by
the phase (probability) function
\begin{equation}
\label{eq:phase}
W(\theta) \propto 1 + \frac{R}{Q}\cos^2\theta,
\end{equation}
where
$\theta$ is the angle between $\mathbf{\hat{n}}_i$ and the outgoing direction
$\mathbf{\hat{n}}_f$, and $R/Q$ is the degree of polarization for $90^\circ$
scattering. For reasons of symmetry, the scattering must always be isotropic in
the azimuthal direction and hence independent of $\phi$.
For scattering in the line center, i.e. for $x < x_{\mathrm{cw}}$,
transitions to the $2P_{1/2}$ state results
in isotropic scattering such that $R/Q = 0$, while the $2P_{3/2}$ transition
causes some polarization, resulting in $R/Q = 3/7$ \citep{ham40}.
Since the spin multiplicity is
$2J + 1$, with $J$ the angular momentum of the state,
the probability of being excited to the $2P_{3/2}$ state is twice as
large as being excited to the $2P_{1/2}$ state\footnote{For the environments
produced here, transitions to the $2S$ state and subsequent
destruction of the photon through two-photon processes can be neglected.}.
For scatterings in the wing, polarization for $\pi/2$ scattering is maximal,
i.e.~$R/Q = 1$ \citep{ste80}.

The transition between the wing and the core is taken to occur at
$x_{\mathrm{cw}}$. Obviously, the phase function does not change abruptly at
this point, but rather varies continuously from one to another in some
fashion. However, the difference in the final outcome is very small, even if
a single phase function is used for all scatterings.

In the observers frame, the final frequency is then
\begin{equation}
\label{eq:xf}
x_f = x_i - u_{||} + \mathbf{\hat{n}}_f \cdot \mathbf{u}
   + g(1 - \mathbf{\hat{n}}_i \cdot \mathbf{\hat{n}}_f),
\end{equation}
where the factor $g = h_{\mathrm{Pl}} \nu_0 / m_{\mathrm{H}} c v_{\mathrm{th}}$
\citep{fie59}, with $h_{\mathrm{Pl}}$ the Planck constant, accounts for the
recoil effect.

After the scattering, the photon is assigned a new, random value of $\tau$ and
continues its journey.

\subsection{Observations}
\label{sec:obs}

Following the above scheme, the photon is trailed as it scatters trough real
and frequency space, until eventually it escapes the computational box.
Subsequently, this procedure is repeated for a number ($n_{\mathrm{ph}}$ in
total) of photons sufficiently
large that the result converges. Here, ``convergence'' is defined as the change
in the desired result after $n_{\mathrm{ph}}$ photons as compared to the result
after $n_{\mathrm{ph}}$ photons being on the $\sim1$\% level.
If we merely concern ourselves with the
characteristic angularly averaged spectrum of escaping Ly$\alpha$ photons,
$n_{\mathrm{ph}}$ needs not be very large, of the order $10^3$.
However, since in general the morphology of a galaxy may very well cause an
anisotropic luminosity, it may be more interesting to see how
the system would appear when observed from a given angle, at a distance given
by the redshift of the galaxy. Because the ratio of photons escaping in a
particular direction is effectively zero, following \citet{yus84} we calculate
instead \emph{for each scattering} and for each photon the probability of
escaping the medium in the direction of the
observer, or, in fact, six different observers situated in the positive and
negative directions of the three principal axes, as
$ W(\theta) e^{-\tau_{\mathrm{esc}}}$,
where $\theta$ is now given by the angle between
$\mathbf{\hat{n}}_i$ and the direction of the observer and
$\tau_{\mathrm{esc}}$ is the optical
depth of the gas lying between the scattering event and the edge of the
computational box (integrated through the intervening cells and of course
taking into account the different bulk velocities of the cells).

This probability is added as a
weight to a three-dimensional array (``CCD'') of two spatial and one spectral
dimension. Each pixel suspends a solid angle $\Omega_{\mathrm{pix}}$
of the computational box.
The total surface brightness
SB$_{\mathrm{pix}}$ of the area covered by the pixel, measured in energy per
unit time, per unit area \emph{at the location of the observer}, per unit solid
angle suspended by the pixel is then
\begin{equation}
\label{eq:SBpix}
\textrm{SB}_{\mathrm{pix}} = \frac{L_{\mathrm{tot}}}
                                  {d_L^2 \Omega_{\mathrm{pix}}}
                             \frac{1}{n_{\mathrm{ph}}}
   \sum_{\mathrm{ph.,scat.}} W(\theta) e^{-\tau_{\mathrm{esc}}},
\end{equation}
where $d_L$ is the luminosity distance given by the redshift, and the sum is
over all photons and all scatterings. Note that
Eq.~(\ref{eq:SBpix}) does not contain a factor $1/4\pi$, due to the fact that
the phase functions are normalized to unity.

Eq.~(\ref{eq:SBpix}) is the SB that an observer would measure
at a distance $d_L$ from the galaxy. Hence, this is the interesting quantity
for comparing with actual observations. Theorists tend to be more concerned
with the intrinsic SB, i.e.~the flux measured by a hypothetical observer at the
location of the source. In this case Eq.~(\ref{eq:SBpix}) must be multiplied
by a factor $(1 + z)^4$, and the SB is then measured in energy per time per
area.

When a sufficient number of photons has been propagated,
the 3D array can be collapsed along the frequential direction to give
a ``bolometric'' Ly$\alpha$ SB map, along the two spatial directions to give
the integrated
spectrum, or along all directions to give the total flux received from the
source. Since in fact we obtain a full spectrum for each pixel, it is also
possible to perform 2D (long slit) spectroscopy, giving frequency as a function
of position of a selected part of the image.

For the spectra to converge, $n_{\mathrm{ph}}$ should be of order $10^4$.
The luminous regions of the SB maps need $n_{\mathrm{ph}} \sim 10^5$ to
converge, while the outer regions need several $10^6$. However, when smoothing
the maps so as to simulate the effect of the atmosphere, or in any case a
finite angular resolution, and averaging
the SB maps in the azimuthal direction to produce SB profiles, less photons are
needed, of order $10^5$. In the simulations described in \S\ref{sec:app},
$n_{\mathrm{ph}} \sim 10^6$--$10^7$ was used.


\section{Testing the Code}
\label{sec:tests}

\subsection{Individual Scatterings}
\label{sec:indiscat}

\begin{figure}
\epsscale{1.1}
\plotone{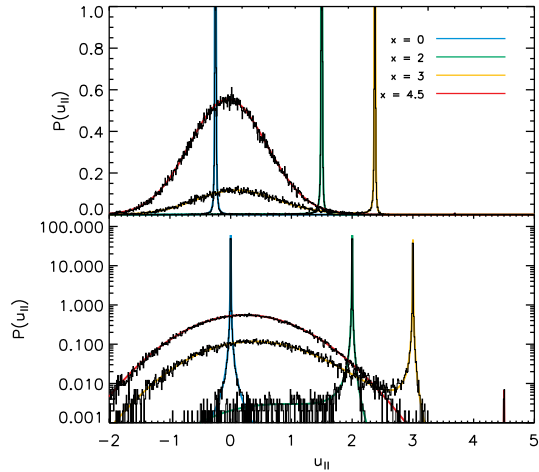}
\caption{Probability distribution $P(u_{||})$ of parallel velocities
         $u_{||}$ of the scattering atom
         for four values $x$ of the frequency of the incoming photon, as
         generated from Eq.~(\ref{eq:fu}). For photons in the line center
         ($x = 0$, \emph{blue}), $P(u_{||})$ resembles
         the natural line broadening function. For successively larger, but
         relatively low frequencies ($x = 2$, \emph{green}, and $x = 3$,
         \emph{yellow}), the photon still has a
         fair chance of being scattered by an atom to which it appears
         close to resonance. For larger frequencies ($x = 4.5$, \emph{red}),
         however, atoms
         with sufficiently large velocities are so rare that the
         distribution instead resembles a regular Gaussian, slightly shifted
         toward $u_{||} = x$. The method for
         generating $u_{||}$ is quite good at resolving the resonance
         peak. This is particularly visible in the logarithmic plot
         (bottom panel).}
\label{fig:uII}
\end{figure}

The various probability distribution generators were tested against their
analytical solutions (in the case such solutions exist; otherwise against
numerical integration). We show here only the result for the parallel
velocities $u_{||}$ (Fig.~\ref{fig:uII}).

\begin{figure*}
\epsscale{0.9}
\plotone{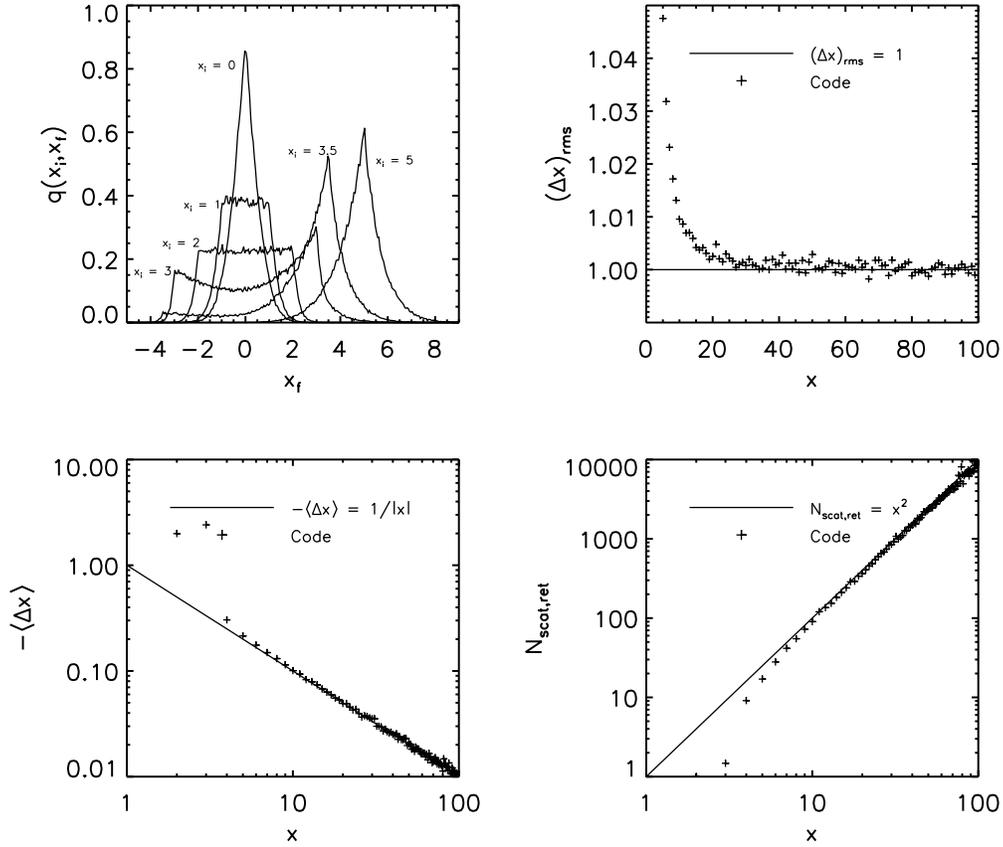}
\caption{Tests of the relation between the frequency $x_i$ of the incoming
         photon, and the frequency $x_f$ of the outgoing photon (top left).
         For photons close to the line center, frequencies are distributed more
         or less uniformly over the line profile. For larger $x$, frequencies
         close to the incoming frequency are preferred, but also frequencies
         of opposite sign.
         The distribution follows that predicted by \citet{hum62}.
         For even
         larger frequencies, photons are less likely to be scattered by atoms
         to which they are at resonance, and the outgoing frequency is then
         only a few Doppler widths away from the ingoing. For sufficiently
         large $x$, the rms shift $(\Delta x)_{\mathrm{rms}} \rightarrow 1$
         (top right), the mean shift
         $\langle\Delta x \rangle \rightarrow -1/|x|$ (bottom left), and the
         average number of scattering needed to return to the core
         $N_{\mathrm{scat,ret.}} \rightarrow x^2$ (bottom right), as predicted
         by \citet{ost62}.}
\label{fig:ost}
\end{figure*}

To test the individual scatterings, Fig.~\ref{fig:ost} shows the relation
between the frequency of the incident and of the scattered photon, compared
with the exact redistribution function as formulated by \citet{hum62}.
Furthermore, the rms and mean shift, and the average number of scattering
before returning to the core for wing photons are shown.
For $x \rightarrow \infty$, the values
are seen to converge to the results derived by \citet{ost62}, and given by
Eqs.~(\ref{eq:rms}) and (\ref{eq:meanx}).

\subsection{Uniform Slab of Gas}
\label{sec:neutest}

\begin{figure}
\epsscale{1.0}
\plotone{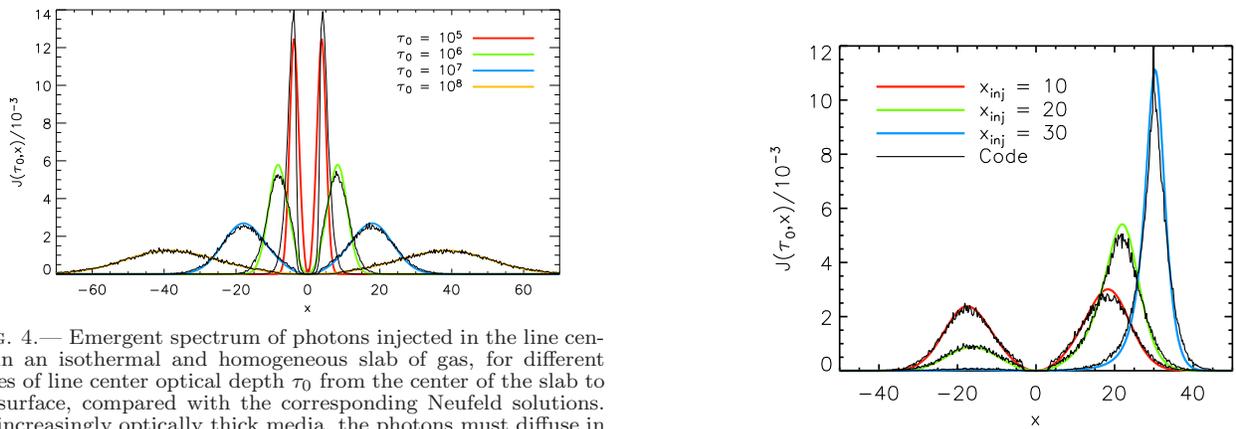}
\caption{Emergent spectrum of photons injected in the line center in an
         isothermal and homogeneous slab of gas, for different values
         of line center optical depth $\tau_0$ from the center of the slab to
         the surface, compared with the
         corresponding Neufeld solutions. For increasingly optically
         thick media, the photons must diffuse in frequency further and
         further from the line center in order to escape the medium.
         For all simulations, $T = 10^4$ K (corresponding to $a = 0.00047$) and
         $n_{\mathrm{ph}} = 10^5$
         was used. The analytical solution becomes increasingly more
         accurate as $\tau_0 \to \infty$.}
\label{fig:neufeld}
\end{figure}

\begin{figure}
\epsscale{1.0}
\plotone{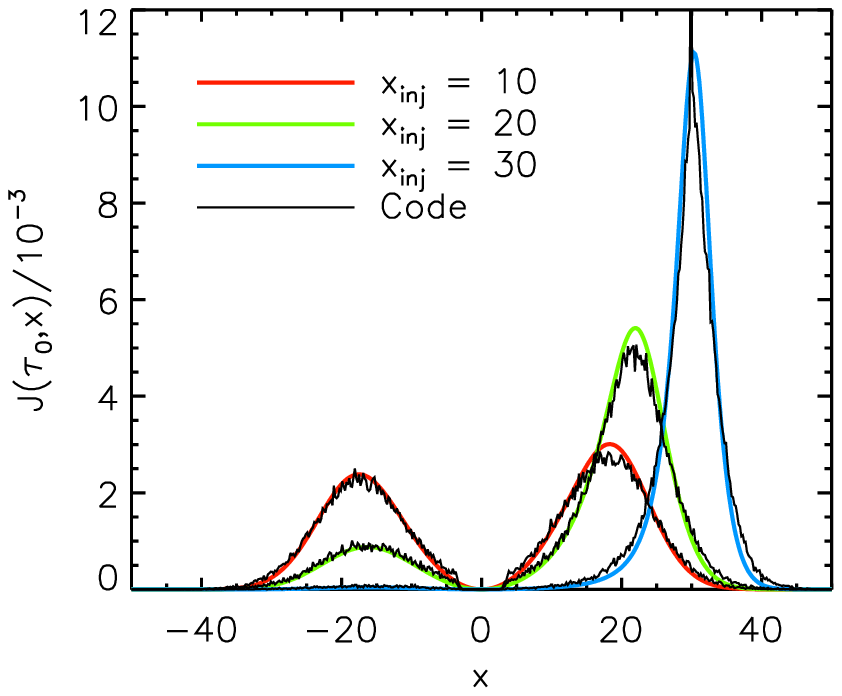}
\caption{Emergent spectrum of $10^5$ photons injected with different
         initial frequencies $x_{\mathrm{inj}}$ in a slab of line
         center optical depth $\tau_0 = 10^7$ and temperature
         $T = 10^4$ K (corresponding to $a\tau_0 = 4700$), compared with the
         corresponding Neufeld solutions.}
\label{fig:x_inj}
\end{figure}

The most basic confirmation of the reliability of the code is a test of the
Neufeld solution. Hence, a simulation of a slab (i.e.~with the $x$- and
$y$-dimension set to infinity) is run in which the bulk velocity of the
elements is set to zero, while the temperature and hydrogen density are
constant in such a way as to give the desired line center optical depth
$\tau_0$ from the center of the slab to the surface. Base cells are refined
in arbitrary locations, to an arbitrary level of refinement. Also,
$W(\theta) =$ constant is used and the recoil term in Eq.~(\ref{eq:xf}) is
omitted to match the assumptions made by Neufeld. The result for different
values of $\tau_0$ is shown in Fig.~\ref{fig:neufeld}, while the result of
varying the initial frequency is shown in Fig.~\ref{fig:x_inj}.
For the lowest
optical depth ($\tau_0 = 10^5$, corresponding to $a\tau_0 = 47$ at $T = 10^4$
K), the fit is not very accurate. However, this is not an artifact caused by,
e.g., a too low number of photons in the simulation, but merely reflects the
fact that the Neufeld solution is no longer valid when the optical depth becomes
too low (at low optical depths, the transfer of photons is no longer dominated
by wing scatterings, where the line profile can be approximated by a power law).

\begin{figure}
\epsscale{1.0}
\plotone{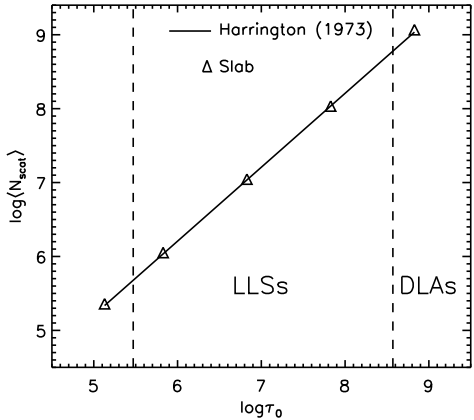}
\caption{Average number of scatterings $\langle N_{\mathrm{scat}} \rangle$
         (triangles) for
         different line center optical depths $\tau_0$, compared with
         the analytical solution (red) given by Eq.~(\ref{eq:Nscat}).
         The dashed
         lines indicate the regions of optical depths for Lyman-limit
         systems (LLSs) and damped Lyman alpha systems (DLAs). A temperature
         of $T = 10$ K was used for all simulations, and the number of photons
         per simulation varied from $10^3$ to $\sim10^5$ for the highest
         and lowest optical depths, respectively.}
\label{fig:Nscat}
\end{figure}

Figure \ref{fig:Nscat} shows the average number of scatterings
$\langle N_{\mathrm{scat}} \rangle$.
%
Of course, in this case
a non-accelerated version of the code (i.e.~$x_{\mathrm{crit}} = 0$) was used,
since we are interested in the true number of scatterings. To get a feeling for
the physical significance of the optical depths, the region of
$\tau_0$ is divided into the domains of the so-called Lyman-limit systems
(LLSs) and damped Ly$\alpha$ systems (DLAs), characterized by limiting neutral
hydrogen column densities of
$N_{\textrm{{\scriptsize H}{\tiny \hspace{.1mm}I}}} = 10^{17.2}$ cm$^{-2}$
and
$N_{\textrm{{\scriptsize H}{\tiny \hspace{.1mm}I}}} = 10^{20.3}$ cm$^{-2}$,
respectively.

\subsection{Gas Bulk Motion}
\label{sec:motion}

Except for slightly different factors, \citet{dij06} found that the emergent
spectrum, its maximum, and
the average number of scatterings of an isothermal, homogeneous sphere with no
bulk velocity of the gas resemble those of the slab. Although not shown here,
such simulations were also carried out, and were run in two different ways;
with a version of the code that has
concentric shells instead of cells, thus exploiting fully the spherical
symmetry, and a normal, cell-based version, the output of which converges to
the former for sufficiently high resolution.

To test if the implementation of the bulk velocity scheme produces reliable
results, we inspect the emergent spectrum of a sphere subjected to isotropic,
homologous expansion or collapse. Thus, the velocity
$\mathbf{v}_{\mathrm{bulk}}(\mathbf{r})$ of a
fluid element at a distance $\mathbf{r}$ from the center is set to
\begin{equation}
\label{eq:vr}
\mathbf{v}_{\mathrm{bulk}}(\mathbf{r}) = \mathcal{H} \mathbf{r},
\end{equation}
where the Hubble-like parameter $\mathcal{H}$ is fixed such that the velocity
increases linearly from 0 in the center to a maximal absolute velocity
$v_{\mathrm{max}}$ at the edge of the sphere ($r = R$):
\begin{equation}
\label{eq:vmax}
\mathcal{H} = \frac{v_{\mathrm{max}}}{R},
\end{equation}
with $v_{\mathrm{max}}$ positive (negative) for an expanding (collapsing)
sphere.

\begin{figure*}
\epsscale{1.0}
\plotone{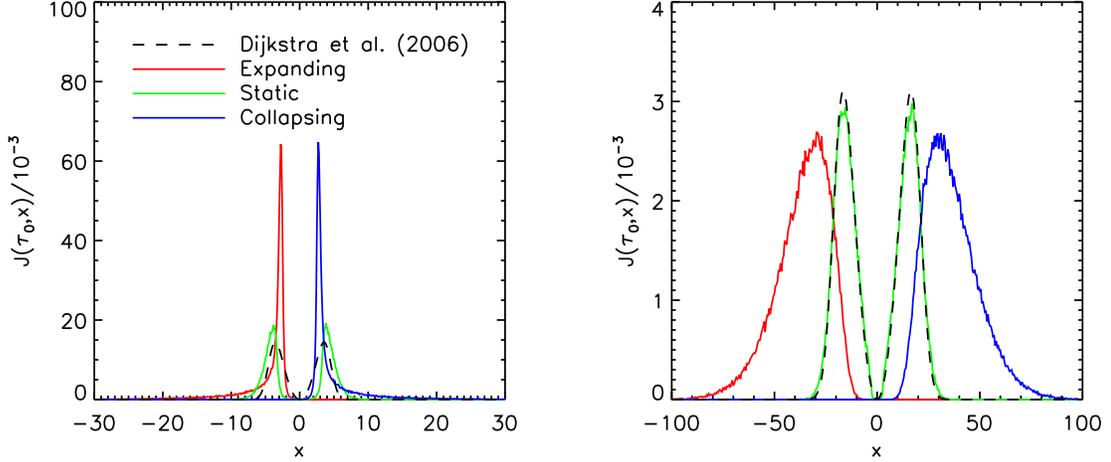}
\caption{Emergent spectrum from an isothermal ($T = 10^4$ K) and
         homogeneous sphere of gas undergoing isotropic expansion (red)
         or contraction (blue) in such a way that the velocity at the
         edge of the sphere is $v_{\mathrm{max}} \pm200$ km s$^{-1}$.
         Left panel shows the result for a column density
         $N_{\textrm{{\scriptsize H}{\tiny \hspace{.1mm}I}}}$ from the
         center to the edge of $2\times10^{18}$ cm$^{-2}$,
         corresponding to $\tau_0 = 1.2\times10^5$ and characteristic
         of a typical LLS. Right panel shows the result for
         $N_{\textrm{{\scriptsize H}{\tiny \hspace{.1mm}I}}} =
         2\times10^{20}$ cm$^{-2}$ ($\tau_0 = 1.2\times10^7$),
         characteristic of a typical DLA. Also shown is the result from
         a simulation with $v_{\mathrm{bulk}} = 0$ (black dashed), and the
         analytical solution for the static sphere (green) as given
         by \citet{dij06}. For the LLS, $\tau_0$ is clearly
         too small to give an accurate fit.}
\label{fig:LLSDLA}
\end{figure*}

For $T \neq 0$ K, no analytical solution for the spectrum exists.
Qualitatively, we expect an expansion to cause a suppression of
the blue wing and an enhancement of the red wing of the spectrum. The reason
for this is that photons blueward of the line center that
would otherwise escape the medium, are shifted into resonance in the reference
frame of atom lying closer to the edge, while red photons escape even more
easily. Conversely, a collapsing sphere will exhibit an enhanced blue wing and
a suppressed red wing. This is indeed seen in Fig.~\ref{fig:LLSDLA}.
Another
way to interpret this effect is that photons escaping an expanding cloud are,
on the average, doing work on the gas, thus losing energy, and vice versa for
a collapsing cloud.

\begin{figure}
\epsscale{1.0}
\plotone{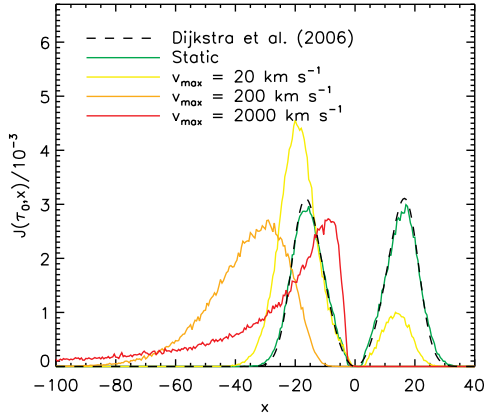}
\caption{Emergent spectrum from an isothermal ($T = 10^4$ K) and
         homogeneous sphere of hydrogen column density
         $N_{\textrm{{\scriptsize H}{\tiny \hspace{.1mm}I}}} =
         2\times10^{20}$ cm$^{-2}$ (a DLA) undergoing isotropic
         expansion with different maximal velocities $v_{\mathrm{max}}$
         at the edge of the sphere. For increasing $v_{\mathrm{max}}$,
         the peak of the profile is pushed further away from the line
         center. However, if $v_{\mathrm{max}}$ becomes too large, the
         medium becomes optically thin and the peak moves back toward
         the center again.}
\label{fig:expand}
\end{figure}

In Fig.~\ref{fig:expand}, results for a sphere of gas expanding at different
velocities are shown.
For increasing $v_{\mathrm{max}}$, the position of the
red peak is progressively enhanced and displaced redward of the line center.
However, above a certain threshold value the velocity gradient becomes so large
as to render the medium optically thin and allow less redshifted photons to
escape, making the peak move back toward the line center again.

The results matches closely those found by previous authors
\citep{zhe02,tas06a,ver06}.

\section{Semianalytical Acceleration Scheme}
\label{sec:acc}

Most of the computing time is spent in the very dense cells. Since each
cell is in fact a ``uniform'' cube,
i.e.~a cube of homogeneous and isothermal gas,
if an analytical Neufeld-equivalent solution for the distribution of frequency
exists, it would be possible to skip a great number of scatterings and thus
speed up the code further.

The slab solution is an alternate series which can be written in closed form.
Unfortunately, this is not feasible for the cube solution, but under certain
approximations, \citet{tas06b} found that it is still possible to write it as
an alternate series. The problem is that, whereas for the slab the terms
quickly die off, the same is not true for the cube. In fact she found that to
achieve an accuracy better than 3\%, one must exceed 30 terms.

Hence, it seems more convenient to seek a ``Neufeld-based'' approximation.
Since for the cube, the radiation can escape from six faces rather than just
two, we may expect the emergent radiation to be described by a function similar
to the slab solution, but using a lower value of $a\tau_0$.

\subsection{Emergent Spectrum}
\label{sec:x_cube}

\begin{figure}
\epsscale{1.0}
\plotone{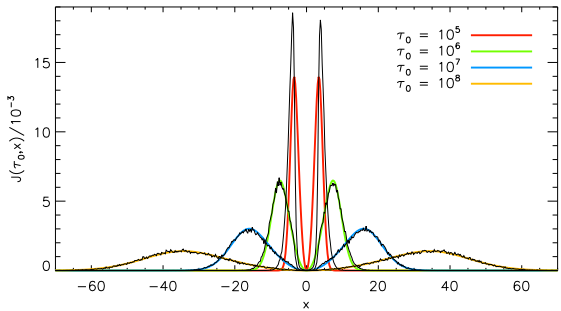}
\caption{Emergent spectra of a uniform cube of different optical
         depths. Neufeld profiles are fitted to the spectra using
         $\eta a\tau_0$, with $\eta = 0.71$
         for all $\tau_0$.}
\label{fig:x_cube}
\end{figure}

Toward these ends, a series of simulations is run in which photons are emitted
isotropically from the center of a cube of constant --- but different ---
temperature and density,
and zero bulk velocity. The distance from the center to each face is $z_0$. We
will investigate optical depths $\tau_0 = 10^5$, $10^6$, $10^7$, and $10^8$
(measured along the shortest path from center to face). In
all simulations, $n_{\mathrm{ph}} = 10^5$, and different temperatures are
tested. We then fit a Neufeld profile to the emergent spectrum, using
$\eta a \tau_0$
as the independent variable, where $\eta$ is the parameter to be determined.
A priori, we have no reason to believe that the same value of $\eta$, if any,
should be able to describe all optical depths. However, it is found that, save
for the lowest optical depth ($\tau_0 \sim 10^5$),
excellent fits are obtained using
\begin{equation}
\label{eq:eta}
\eta = 0.71.
\end{equation}
This is seen in Fig.~\ref{fig:x_cube}.
\ \\

\subsection{Directionality of the Emergent Photons}
\label{sec:dir_cube}

In realistic,
cosmological simulations, the direction with which the photons exit
the cell is also important. Since in the limit $\tau_0 \to \infty$, any finite
size step \emph{not} perpendicular to the surface will just shift to position
of the photons in the parallel direction, for extremely optically thick slabs,
the photons should have a
tendency to exit perpendicular to the surface. In this case, \citet{phi86}
found that the directionality of the emergent radiation approaches that of
Thomson scattered radiation from electrons, with
intensity
\begin{equation}
\label{eq:ImuI0}
\frac{I(\mu)}{I(0)} = \frac{1}{3} \left( 1 + 2\mu  \right),
\end{equation}
where $\mu = \cos\theta$, with $\theta$ the angle between the outgoing
direction $\hat{\mathbf{n}}_f$ of the photon and the normal to the surface.

Since the number of photons emerging at $\mu$ is $\propto I(\mu)\mu\,d\mu$, the
probability $P(\le\mu)$ of exiting the slab with $\mu \le \mu'$ is
\citep{tas06b}
\begin{eqnarray}
\label{eq:Pltmu}
\nonumber
P(\le\mu) & = &\frac{\int_0^{\mu'} (1+2\mu)\mu\,d\mu}
                    {\int_0^1      (1+2\mu)\mu\,d\mu}\\
          & = & \frac{\mu'^2}{7} \left( 3 + 4\mu' \right).
\end{eqnarray}

\begin{figure*}
\epsscale{0.9}
\plotone{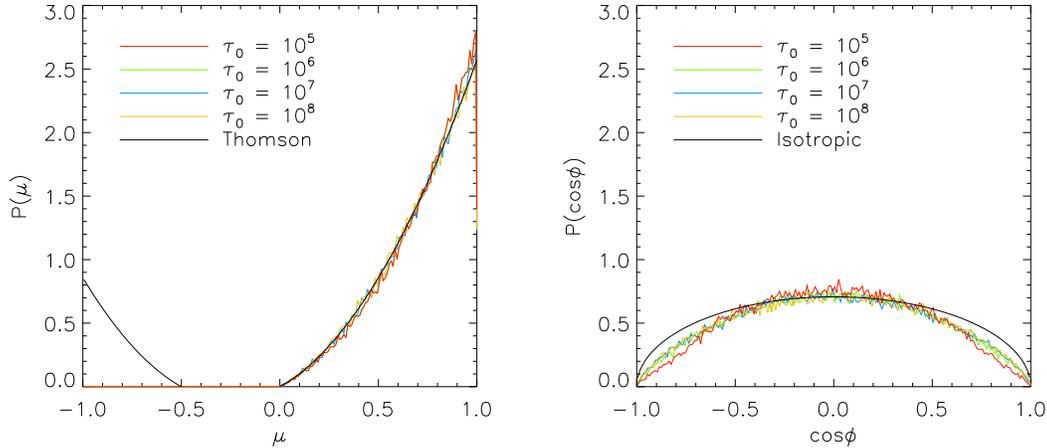}
\caption{Directionality of the photons emerging from a uniform cube
         for different values of $\tau_0$. In
         the direction perpendicular to the face of the cube (\emph{left}),
         $\hat{\mathbf{n}}_f$ follows the distribution given by
         Eq.~(\ref{eq:Pmucube}), while in the azimuthal direction
         (\emph{right}) there is a slight deviation from isotropy.}
\label{fig:mu_cube}
\end{figure*}

We confirm that this is also an excellent description for a cube
(Fig.~\ref{fig:mu_cube}).

The probability distribution is found by differentiating Eq.~(\ref{eq:Pltmu})
and recognizing that $\mu$ must be positive for the photon to escape:
\begin{equation}
\label{eq:Pmucube}
P(\mu) =  \left\{ \begin{array}{ll}
           \frac{6}{7} (\mu + 2\mu^2) & \textrm{for 0 $< \mu \le$ 1}\\
           0                          & \textrm{otherwise},
\end{array}
\right.
\end{equation}

Since Eq.~(\ref{eq:Pmucube}) is valid for all six faces of the cube, the
azimuthal angle $\phi$ parallel to the face cannot, as in the case of a slab,
be evenly distributed in $[0,2\pi]$ \citep{tas06b}. However, as can be seen
from Fig.~\ref{fig:mu_cube}, the deviation from uniformity is quite small, and
can probably be neglected. Furthermore, it seems less pronounced, the higher
the optical depth.

\subsection{Point of Escape}
\label{sec:exitcube}

\begin{figure}
\epsscale{1.0}
\plotone{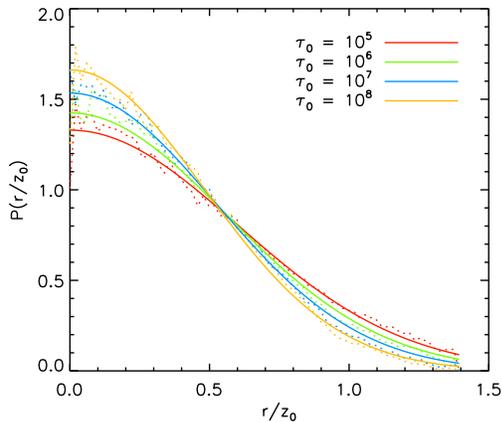}
\caption{Probability distribution (solid lines) of the exiting point for
         photons emerging from a uniform cube of side length $2 z_0$, as a
         function of distance $r$ from the center of the face,
         normalized to $z_0$ , for different values of $\tau_0$.
         The distributions have been calculated as best fits to the
         corresponding simulated surface brightness
         profiles (dotted lines), as given by Eq.~(\ref{eq:SBcube}).}
\label{fig:SBcube}
\end{figure}

The final parameter characterizing the photons escaping the cube is the point
$\mathbf{x}_{\mathrm{esc}}$ where it crosses the face. Figure \ref{fig:SBcube}
shows the azimuthally averaged SB profiles of the emergent radiation as a
function of distance from the center of the face, for different optical depths.
It is found that the SB profile is fairly well described by
a truncated Gaussian
\begin{equation}
\label{eq:SBcube}
\textrm{SB}(r/z_0) =  \left\{ \begin{array}{ll}
 \frac{2}{\sqrt{2\pi}\sigma_{\mathrm{SB}}}
 e^{-(r/z_0)^2/2\sigma_{\mathrm{SB}}^2}& \textrm{for $0\le r\le z_0\sqrt{2}$}\\
           0                           & \textrm{for $r > z_0\sqrt{2}$}.
\end{array}
\right.
\end{equation}
%
The dispersion $\sigma_{\mathrm{SB}}$ of the SB decreases very slowly with
optical depth, and can be written as
$\sigma_{\mathrm{SB}} = 0.48 - 0.04 \log \tau_0/10^8$.
However, in the context of a cell-based structure, one might state that it is
meaningless to discuss differences in position on scales smaller than the size
of a cell, and it is found that final results are not altered by simply setting
$\sigma_{\mathrm{SB}} = 0.5$.
\ \\

\subsection{Implementation of the Cube Solution}
\label{sec:impl}

With the probability distributions of frequency, direction and position for the
photons escaping the cell, we are now able to accelerate the code further:
every time a photon finds itself in a host cell of $a\tau_0$ higher than some
given threshold, which to be conservative we define as
$a\tau_0 \gtrsim 2\times10^3$, an \emph{effective cell} with the photon in the
center is built, with ``radius'' $z_0$ equal to the distance from the photon to
the nearest face of the host cell.
Since the effective cell is always completely circumscribed by the host cell,
its physical parameters are equal to those of its host cell.

If the value of $a\tau_0$ in the effective cell, $(a\tau_0)_{\mathrm{eff}}$, is
below the threshold (i.e.~if the photon is too close to the face
of the host cell), the normal scheme is used. Otherwise, the photon is assigned
a new frequency according to the effective Neufeld distribution:
drawing a univariate
$\mathcal{R}$ and setting this equal to the Neufeld-equivalent cube
solution\footnote{Of course normalized to unity instead of the usual
$1/4\pi.$} integrated from $-\infty$ to $x$ yields (after some algebra)
\begin{eqnarray}
\label{eq:intJcube}
\nonumber
\mathcal{R} & = & \int_{-\infty}^{x_f} J_{\mathrm{cube}}(\tau_0,x)dx\\
            & = & \frac{2}{\pi} \tan^{-1}
                  e^{\sqrt{\pi^3/54}
                     (x_f^3-x_i^3) / 
                     \eta (a\tau_0)_{\mathrm{eff}}}.
\end{eqnarray}
Inverting the above expression, the frequency $x$ of the photon then becomes
\begin{equation}
\label{eq:x_cube}
x_f = \left( \sqrt{\frac{54}{\pi^3}} \eta (a\tau_0)_{\mathrm{eff}}
    \ln \tan \frac{\pi\mathcal{R}}{2}  + x_i^3 \right)^{1/3}.
\end{equation}

Finally, the direction and the position of the photon is determined from
Eqs.~(\ref{eq:Pmucube}) and (\ref{eq:SBcube}) (with a probability of escaping
from a given face equal to $1/6$), whereafter it continues its journey.

\section{Application to Cosmological Simulations}
\label{sec:app}
To demonstrate the potency of the developed code, we apply it to a number of
different simulated galaxies.
Specifically, we will study three young ``Lyman-break galaxies'' (LBGs),
for the purpose of the present study dubbed K15, K33, and S115, at a
redshift of $z = 3.6$, at which time the Universe was $\sim 1.8$ Gyr old. These
three galaxies are representative of typical galaxies in the sense
that the two larger evolve into Milky Way/M31-like disk galaxies at $z = 0$,
whereas S115 becomes a somewhat smaller disk galaxy.
The characteristic circular velocities of the galaxies
at $z = 0$ are $V_{\mathrm{c}} =$ 245, 180 og 125 km s$^{-1}$, respectively.

The cosmological simulation is conducted using an N-body/hydrodynamical
TreeSPH code. A ``pseudo''-RT scheme of the ionizing UV
radiation is accomplished on-the-fly, while a more comprehensive
UV RT scheme is implemented post-process. This is described below.
\ \\

\subsection{Underlying Cosmological Simulations}
\label{sec:undsim}

The numerical simulations are first carried out at low resolution,
but in a large volume of space. Subsequently, interesting galaxy forming
regions are resimulated at high resolution.
Typically, resimulations are performed at 64$\times$ higher mass resolution.
In order to
check the significance of the resolution, also lower resolution (8$\times$)
and ultrahigh resolution (512$\times$) simulations are executed.

The simulations are started at an
initial redshift $z_i = 39$, at which time there is only dark matter (DM) and
gas SPH particles. The latter eventually evolves partly into star particles,
while, in turn, star particles can become gas particles again.
The star formation criteria is described in \citet{som03}.
For K15 and K33, a \citet{kro98} initial mass function (IMF) has been assumed,
while for S115, a \citet{sal55} IMF was used.
A standard $\Lambda$CDM is assumed,
i.e.~$\Omega_m = 0.3$, $\Omega_\Lambda = 0.7$, and the rms linear density
fluctuation on scales of $8 h^{-1}$ Mpc is $\sigma_8 = 0.9$.
In addition to hydrogen and helium, the code also
follows the chemical evolution of C, N, O, Mg, Si, S, Ca, and Fe, using the
method of \citet{lia02}.

Additional information about the simulations can be found in
Tab.~\ref{tab:data}, while Tab.~\ref{tab:physdata} summarizes the physical
properties of the resulting galaxies, demonstrating that they are typical of
galaxies at $z = 3$--5.
Having an $R$-band magnitude of $R \simeq 25$, K15 and K33 could be detected
as LBGs in a survey like the one of
\citet[][$R_{\mathrm{lim}} = 25.5$]{ste03}.
With $R \simeq 28$, S115 is too faint to be detected as an LBG in
\emph{current} surveys, although the limit is being approached
\citep[e.g.][reaching $R_{\mathrm{lim}} \simeq 27$]{saw05}.
\ifnum\astroph=0
  \begin{deluxetable}{lcccccccc}
\else
  \begin{deluxetable*}{lcccccccc}
\fi
\tablecolumns{9}
\tablewidth{0pc}
\tablecaption{Characteristic quantities of the simulations}
\tablehead{
\colhead{Galaxy} & \multicolumn{2}{c}{K15} & \colhead{} &
                   \multicolumn{2}{c}{K33} & \colhead{} &
                    \multicolumn{2}{c}{S115} \\
\cline{2-3} \cline{5-6} \cline{8-9} \\
\colhead{Resolution}& \colhead{8$\times$}  & \colhead{64$\times$} & 
\colhead{}          & \colhead{8$\times$}  & \colhead{64$\times$} &
\colhead{}          & \colhead{64$\times$} & \colhead{512$\times$}
}
\startdata                                                                                                
$N_{\mathrm{p,tot}}$                                 & $3.0\times10^5$     &  $2.2\times10^6$     & &  $1.5\times10^5$    & $1.2\times10^6$      & &  $2.1\times10^5$     & $1.3\times10^6$       \\
$N_{\mathrm{SPH}}$                                   & $1.4\times10^5$     &  $1.0\times10^6$     & &  $7.1\times10^4$    & $5.5\times10^5$      & &  $1.0\times10^5$     & $6.4\times10^5$       \\
$m_{\mathrm{SPH}}$,$m_{\mathrm{star}}$               & $7.4\times10^5$     &  $9.3\times10^4$     & &  $7.4\times10^5$    & $9.3\times10^4$      & &  $9.3\times10^4$     & $1.1\times10^4$       \\
$m_{\mathrm{DM}}$                                    & $4.2\times10^6$     &  $5.2\times10^5$     & &  $4.2\times10^6$    & $5.2\times10^5$      & &  $5.2\times10^5$     & $6.6\times10^4$       \\
$\epsilon_{\mathrm{SPH}}$,$\epsilon_{\mathrm{star}}$ &      382            &        191           & &      382            &      191             & &       191            &      96               \\
$\epsilon_{\mathrm{DM}}$                             &      680            &        340           & &      680            &      340             & &       340            &      170              \\
$l_{\mathrm{min}}$                                   &       20            &        10            & &       20            &       10             & &        10            &        5              \\
\enddata
\tablecomments{Total number $N_{\mathrm{p,tot}}$ of particles, number
               $N_{\mathrm{SPH}}$ of SPH particles only, masses $m$, gravity
               softening lengths $\epsilon$, and minimum smoothing lengths
               $l_{\mathrm{min}}$ of dark matter (DM), gas (SPH), and star
               particles used in the simulation of the three galaxies K15, K33,
               and S115, for different resolutions. Masses are measured in
               $h^{-1}M_\odot$, distances in $h^{-1}$pc.}
\label{tab:data}
\ifnum\astroph=0
  \end{deluxetable}
\else
  \end{deluxetable*}
\fi
\begin{deluxetable}{lccc}
\tablecolumns{4}
\tablewidth{0pc}
\tablecaption{Physical properties of the simulated galaxies}
\tablehead{
\colhead{Galaxy}                             & K15                & K33                & S115 \\
}
\startdata                                                                                               
SFR/$M_\odot$ yr$^{-1}$                      & 16                 & 13                 & 0.5                \\
$M_*/M_\odot$                                & $1.3\times10^{10}$ & $6.5\times10^{9}$  & $2.5\times10^{8}$  \\
$V_{\mathrm{c}}(z=0)$/km s$^{-1}$            & 245                & 180                & 125                \\
$r_{\mathrm{vir}}$/kpc                       & 47.5               & 37.3               & 22.0               \\
$L_{\mathrm{Ly}\alpha}$/erg s$^{-1}$         & $3.3\times10^{43}$ & $1.4\times10^{43}$ & $7.0\times10^{41}$ \\
$L_{\nu,\mathrm{UV}}$/erg s$^{-1}$ Hz$^{-1}$ & $6.7\times10^{28}$ & $5.5\times10^{28}$ & $3.6\times10^{27}$ \\
\enddata
\tablecomments{Star formation rates (SFRs), stellar masses ($M_*$) circular
               velocites ($V_{\mathrm{c}}$), virial radii
               ($r_{\mathrm{vir}}$), Ly$\alpha$ luminosities
               ($L_{\mathrm{Ly}\alpha}$), and UV luminosities
               ($L_{\nu,\mathrm{UV}}$) for the three simulated galaxies K15,
               K33, and S115. All quoted values correspond to a redshift of
               $z = 3.6$, except $V_{\mathrm{c}}$ which is given for $z = 0$.}
\label{tab:physdata}
\end{deluxetable}

The Ly$\alpha$ emission is produced by three different processes
\citep[see also][]{lau07},
viz.~from recombinations in photoionized regions around massive stars
(responsible for $\sim90$\% of the total Ly$\alpha$ luminosity),
gravitational cooling ($\sim10$\%), and a metagalactic UV background (UVB)
photoionizing the external parts of the galaxy ($\sim1$\%).
In the first case, the luminosity is
determined following \citet{far01}, using the code Starburst99 \citep{lei99} to
yield the Lyman continuum (LyC), and assuming a mean LyC photon energy of 1.4
Rydberg and that 0.68 Ly$\alpha$ photons are emitted per photonionization.
The gravitational cooling is accounted for by keeping track of temperature and
ionization state of the gas.
The UVB field is assumed to be that given by \citet{haa96}, where the
gas is treated as optically thin to the UV radiation until the mean free path
of a UV photon at the Lyman limit becomes less than 1 kpc,
at which point the gas is treated as
optically thick and the UV field is ``switched off''.

For a more thorough description of the code, the reader is referred to
\citet{som03}, and to \citet{som06} for recent updates.

\subsection{Ionizing UV Radiative Transfer}
\label{sec:uvrt}

To model the propagation of ionization fronts realistically,
\citet{raz06,raz07}
implemented the following RT scheme as a post-process to the cosmological
simulation: the position of the SPH
particles and their associated physical properties are interpolated from the
50 nearest neighboring particles onto a grid of base resolution $128^3$ cells.
Dense cells are subdivided in eight cells, which are further refined
until no cell contains more than ten particles. 



Around each stellar source,
a system of $12\times4^{n-1}$ radial rays is built that split either as one
moves farther away from the source or as a refined cell is entered,
and $n = 1,2,\ldots$ is the angular resolution level. Once a
radial ray is refined angularly, it stays refined at larger distances
from the source, even when leaving the high-resolution region. In each
cell, the photoreaction number and energy rates due to
photons traveling along ray segments passing through that cell are accumulated.
These rates are then used to update temperature and the ionization state of
hydrogen and helium, which in turn are used to calculate the LyC
opacities used in the RT. In addition
to stellar photons, we also account for ionization and heating by LyC
photons originating outside the computational volume with the FTTE
scheme \citep{raz05} assuming the Haardt-Madau UVB.

Although the UV RT is not coupled to hydrodynamics, hydrodynamical
(shock) heating needs to be taken into account during the RT.
All cells with temperatures above $T_{\rm cr}=3\times10^4{\rm\,K}$ in
the SPH output are considered to be shock-heated, and during the RT only their
ionization state, not their temperature, is updated. For all cells with
temperatures below $T_{\rm cr}$ a hydro-heating term is computed, which
is defined as the amount of heating needed to keep the temperature of
that cell constant if its ionization state stayed at the original
level. This hydro-heating term is used to update both temperature and
ionization of all $T<T_{\rm cr}$ cells. For these cells, a
temperature ceiling of $T_{\rm cr}$ is used to avoid unphysical overheating,
as during the RT calculation heated gas is not allowed to expand.

\subsection{Ly$\alpha$ Radiative Transfer}
\label{sec:lyart}

For the purpose of the Ly$\alpha$ RT, the same AMR grid is used as for the UV
RT. A (250 kpc)$^3$ box is cut out from each $z = 3.6$ snapshot and
is refined to $\mathcal{L}_{\mathrm{max}} = 7$ or 8 levels,           
where $\mathcal{L} = 0$ corresponds to the unrefined (base) grid. Thus, the
cell size of the smallest cells is merely $\sim10$ pc, more than
four orders of magnitude smaller than the computational box itself and comparable
to the size of molecular clouds.

\subsubsection{Significance of the Improved UV RT}
\label{sec:siguvrt}

\begin{figure*}
\plotone{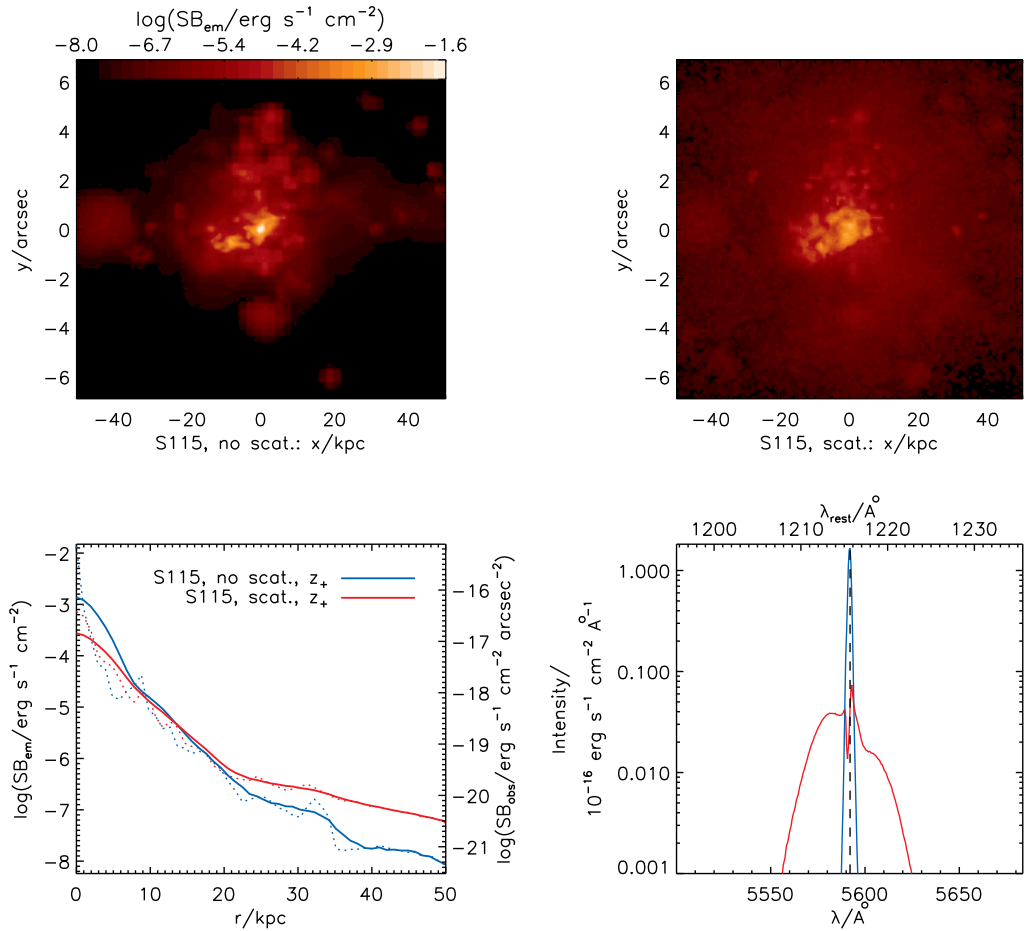}
\caption{Results for the galaxy S115 from the Ly$\alpha$ scattering code
         {\sc MoCaLaTA}. \emph{Top left} plot shows the Ly$\alpha$ surface
         brightness (SB) map of the emitted radiation in the positive
         $z$-direction, i.e. how the galaxy would look if the
         photons did not scatter. Almost all the photons are emitted in the
         center, and almost none are emitted further away than 20 kpc from the
         center. Taking into account scattering (\emph{top right}), the emission
         is clearly much more extended, while the maximum SB is decreased. This
         is also seen in the SB profile (\emph{bottom left}), i.e. the
         azimuthally averaged SB map. Both the true (\emph{dotted curves}) and
         the profiles of the SB smoothed with a seeing of $0\farcs8$
         (\emph{solid curves}) are shown, for both the emitted (\emph{blue})
         and the scattered (\emph{red}) radiation.
         The photons scatter not only in real, but also in frequency space
         (\emph{bottom right}); while the emitted spectrum (\emph{blue}) is
         close to a delta function, the escaping spectrum (\emph{red}) is
         broadened by many {\AA}ngtr\"oms.
         Moreover, due to the fact that the hydrogen cross-section is so large
         for photons in the line center, the spectrum is split up into two
         peaks.}
\label{fig:scatnoscat}
\end{figure*}

Initially, we will focus on K15, consisting of two rather compact ``disks''
embedded in a more extended, $\sim10$--15 kpc thick, sheet-like structure
composed of nonstar-forming H\,\textsc{i} gas,
taken to constitute the $xy$-plane.
Figure \ref{fig:scatnoscat} shows the general effect of the scattering: the
SB is increased in the outskirts of the galaxy, at the expense
of a decrease in the center, where most of the photons are produced. The effect
on the spectrum is also seen: while the only broadening of the input spectrum
visible is due to the bulk motion of the gas elements emitting the photons (the
natural broadening being much smaller), the scattered spectrum is severely
broadened, diminished by more than an order of magnitude, and split up into two
peaks due to the high opacity of the gas for photons in the line center.

\begin{figure*}
\epsscale{1.1}
\plotone{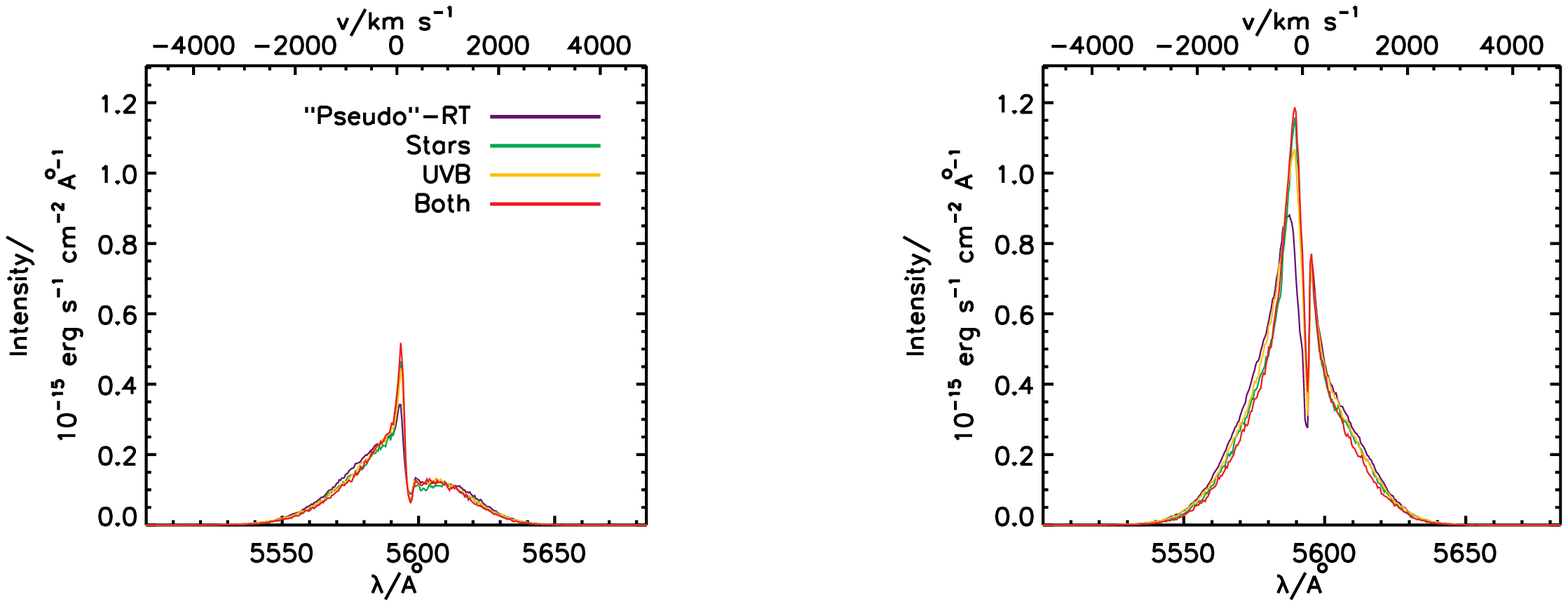}
\caption{Emergent spectrum of the galaxy K15, as seen when observing the
         sheet-like structure in which the galaxy is embedded edge-on
         (\emph{left}) and face-on (\emph{right}).
         \emph{Blue} lines show the
         spectrum for the model without the improved UV RT, while \emph{green},
         \emph{yellow}, and \emph{red} lines show the spectrum when treating
         the UV RT
         properly for the stellar sources only, the UV background only, and
         both, respectively.
         The only real difference is seen in the blue peak of the spectra,
         which is a bit higher for the improved models. In particular, all
         models indicate a moderate net infall of gas, enhancing the
         blue-to-red peak ratio.
         This figure, as well as all following figures, refers to $z = 3.6.$}
\label{fig:AlexSpec}
\end{figure*}

Figure \ref{fig:AlexSpec}
shows the emergent spectrum, as observed when viewing the sheet edge-on and
face-on, respectively. For each case, four spectra are shown:
{\sc i})   using the temperature and ionization from the original SPH run,
{\sc ii})  using the temperature and ionization computed with the LyC transfer
           scheme from Sec.~\ref{sec:uvrt} applied to stellar photons only,
{\sc iii}) using the temperature and ionization from a model in which the UVB
           is traced with the FTTE scheme and there are no stellar photons,
           and
{\sc iv})  using distributions from a model in which LyC RT is computed for
           both stellar and UVB photons.

As expected, fewer photons escape in the plane of the sheet, as the
Ly$\alpha$ optical depth in this direction is greater and more
photons scatter and eventually escape in the directions perpendicular
to the sheet. Note that both edge-on and face-on spectra are sensitive
to the changes in ionization of predominantly low-density regions
which is computed with LyC radiative transfer.
As discussed in \S\ref{sec:neufeld}, the quantity that determines the
shape of the spectrum is the product
$a\tau_0 \propto n_{\textrm{{\scriptsize H}{\tiny \hspace{.1mm}I}}} / T$
(at a fixed physical size).
Generally, a higher temperature will also imply a lower density of neutral
hydrogen, and vice versa, and thus we might expect
$n_{\textrm{{\scriptsize H}{\tiny \hspace{.1mm}I}}} / T$ to change rapidly to
higher or lower values for the improved scheme.
However, for high density cells the change in the ratio
$n_{\textrm{{\scriptsize H}{\tiny \hspace{.1mm}I}}} / T$ is minor when
invoking the improved scheme, in most
cases of order unity. Only in low density cells is this ratio considerably
altered, but since $>90$\% of all scatterings take place in high-density
cells, the overall effect is small. The only notable difference is seen in
the inner part of the spectrum, which is exactly the part that is created by
the low-density regions, since photons near the line center cannot escape from
high-density regions.

\subsubsection{Characteristics of the Emergent Spectrum}
\label{sec:spec}

The double peak profile seen in Fig.~\ref{fig:AlexSpec} is characteristic of
Ly$\alpha$ emission lines;
the high opacity for photons near the line center makes
diffusion to the either side necessary in order to escape the galaxy.
Nonetheless, unlike in previous simple models,
the intensity in the line center is not zero.
The photons that contribute to this intensity are those produced
mainly by gravitational cooling, in the outskirts of the systems.

\begin{figure*}
\epsscale{1.2}
\plotone{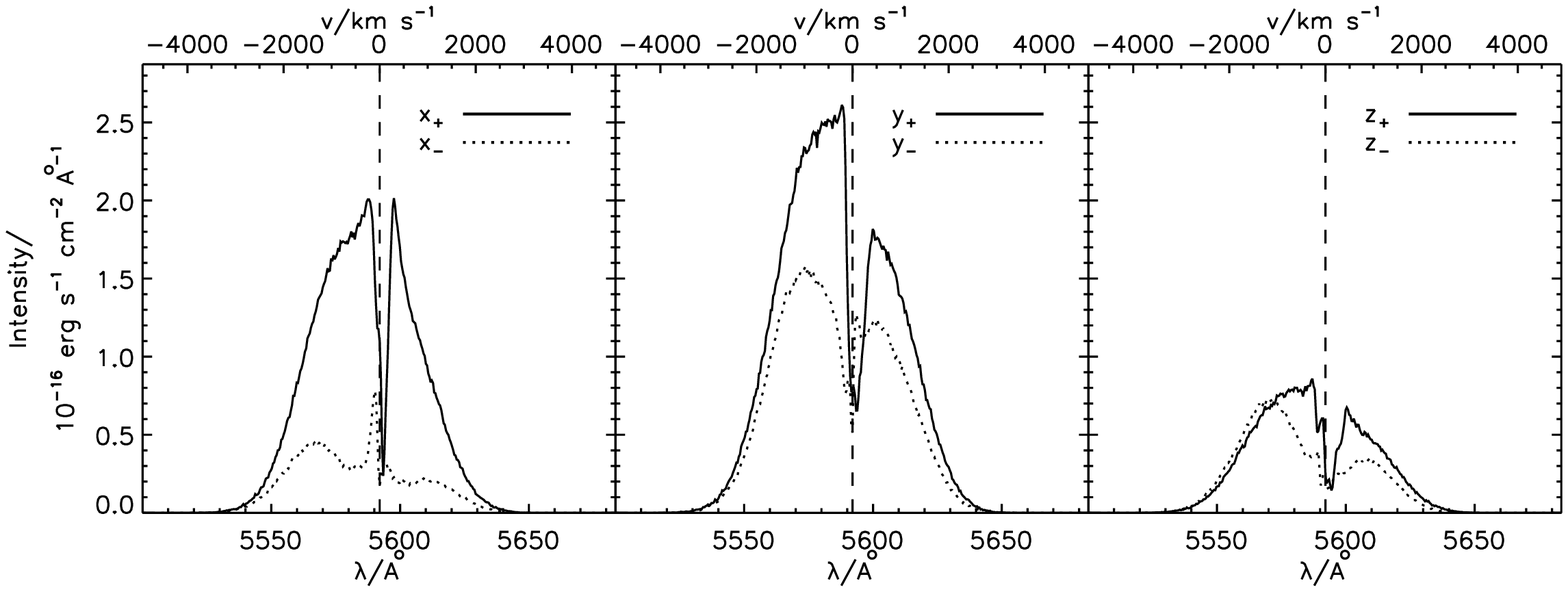}
\caption{Spectral distribution of the photons escaping the galaxy K33 in six
         different directions; along the positive (+) and negative ($-$) $x$-,
         $y$-, and $z$-direction. The dashed line in the middle of each plot
         indicates the line center. The lower abscissa gives the redshifted
         wavelength of the photons while on the upper abscissa, the wavelength
         distance from the line center is translated into recessional velocity.
         The resonant scattering of Ly$\alpha$ is seen to broaden the line by
         several thousands of km s$^{-1}$.}
\label{fig:spec33}
\end{figure*}

\begin{figure*}
\epsscale{1.2}
\plotone{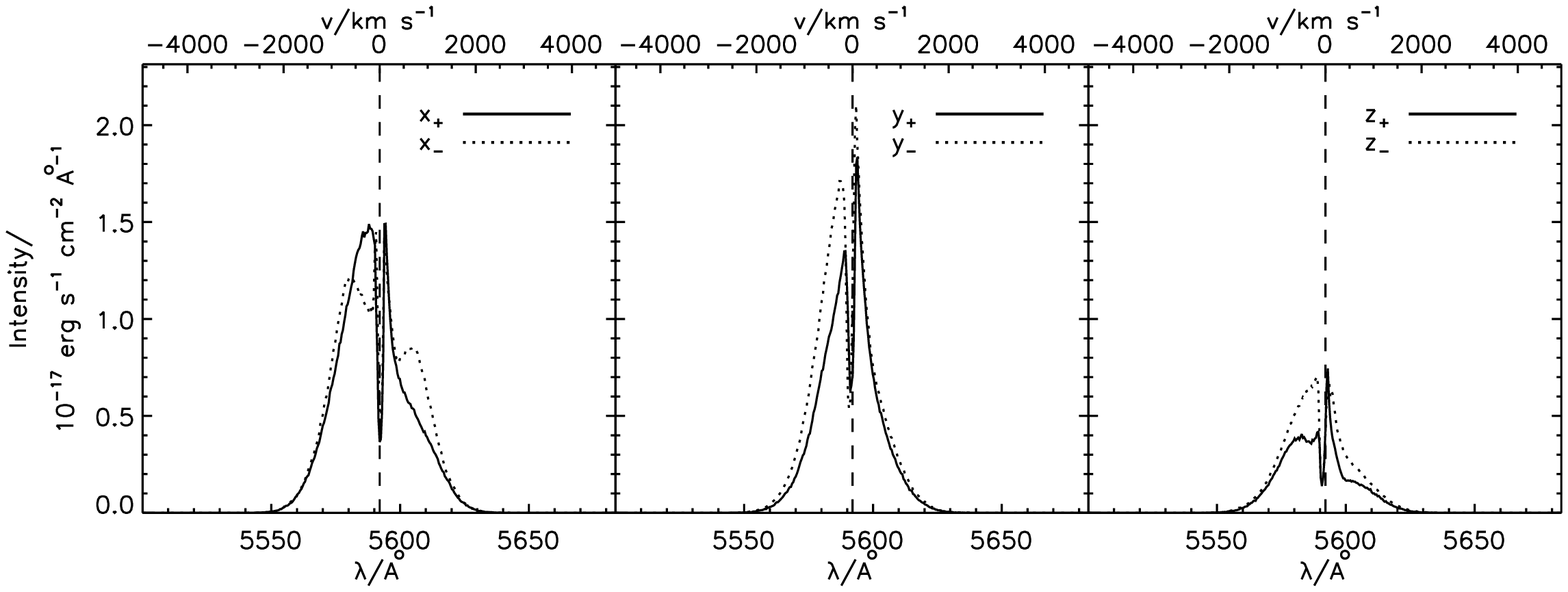}
\caption{Same as Fig.~\ref{fig:spec33}, but for the galaxy S115.}
\label{fig:spec115}
\end{figure*}

Figures \ref{fig:spec33} and \ref{fig:spec115} display the spectra emerging
from galaxies K33 and S115 at $z = 3.6$, in six different directions.
The exact shape varies quite a lot, but all spectra appear to exhibit
the double peak profile. Moreover, they are broadened by several 1000
km s$^{-1}$.

Double peaks have been observed on several occasions
\citep[e.g.,][]{yee91,ven05}.
\citet{tap07}, using a resolution of $R \sim 2000$, found three out of 16
LAEs at redshifts $z \sim 3$--$4$ to have a double-peaked profile,
 while Yamada et al. (priv.~comm.), using $R \sim 1500$ found that
26 of 94 LAEs at redshift 3.1 have the double-peaked profile.
The difference in magnitude of the two peaks can be a signature of
infalling/outflowing gas, cf.~\S\ref{sec:motion}.
In principle, this difference may be used as a probe of the gas dynamics, and
has indeed been used to infer the presence of galactic superwinds.
However, since LAEs are often situated in higher-than-average density regions,
the removal or diminishing of the blue peak might also be caused by
IGM resonant scattering combined with cosmic expansion.

Even if the double peak survives intergalactic transmission, fairly high
resolution is required. With a typical separation $\Delta\lambda$ of the peaks
of the simulated spectra
from a few to $\simeq10$ {\AA}, the resolution must be
$R = 5600 / \Delta\lambda \simeq 500$--$1500$.

It is interesting that while the Ly$\alpha$ profile of the            
sub-Milky Way galaxy K15 also shows
a moderate gas infall, we see some outflow signatures in the                   
spectrum of the lower-mass galaxy S115 from negative $x$- and $y$-directions.
Fitting a Neufeld, or Dijkstra,
profile to the observed spectra can give us an idea of
the intrinsic properties of the system. Unfortunately, due to the
degeneracy between column density and temperature, one would have to gain
knowledge either of the parameters by other means to constrain the other
(e.g.~by inferring column density from the spectrum of a coincident background
quasar, or by assuming a temperature of, say, $10^4$ K, representative of most
of the Ly$\alpha$ emitting gas).
However, clumpiness of the ISM will lower the effective optical depth, making
any inferred value of $a\tau_0$ a lower bound. This is exactly the reason we
need realistic models; galaxies are not isothermal, homogeneous slabs.

\subsubsection{Surface Brightness}
\label{sec:SB}

\begin{figure*}
\epsscale{1.1}
\plotone{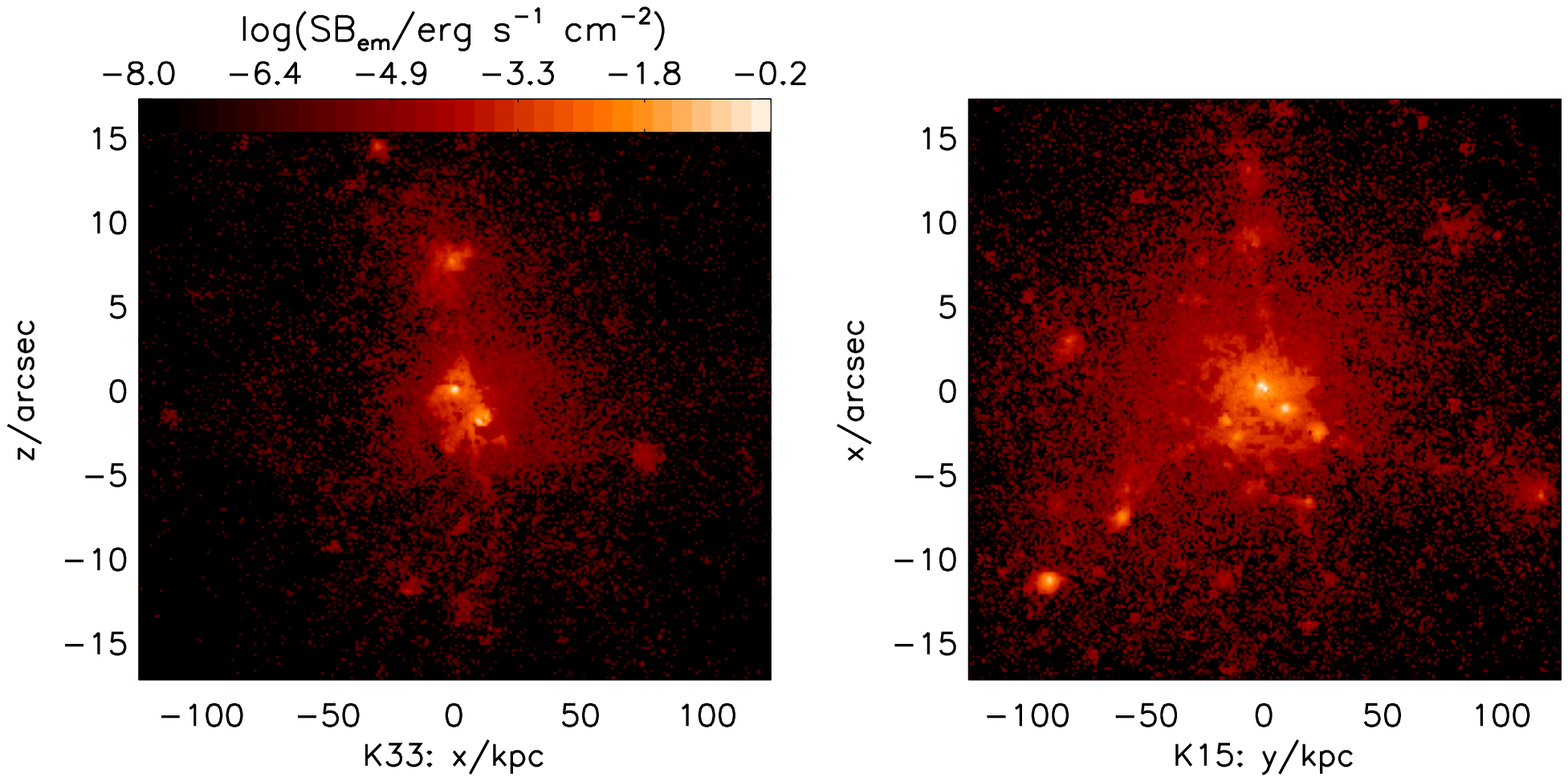}
\caption{Ly$\alpha$ surface brightness maps (integrated over the line) of the
         two galaxies K33 (\emph{left}) and
         K15 (\emph{right}) as viewed from the negative $y$-direction and the
         negative $z$-direction, respectively.}
\label{fig:SBmaps}
\end{figure*}

Figure \ref{fig:SBmaps} displays the Ly$\alpha$ SB maps, integrated over the
line, of two of the galaxies, K15 and K33.
As mentioned already, K15 is embedded in a
sheet-like structure, lying at the intersection of three filaments of gas.
Since the bulk of the photons are produced in the central, star-forming
regions, the total optical depth is larger in the direction parallel
to the sheet than perpendicular to it, and hence we would expect the photons to
escape more easily in the face-on direction.
Similarly, K33 is situated in a filament of
gas, taken to lie along the $z$-axis. Here, we would expect the photons to
escape more easily in the $x$- and $y$-directions.

\begin{figure*}
\epsscale{1.2}
\plotone{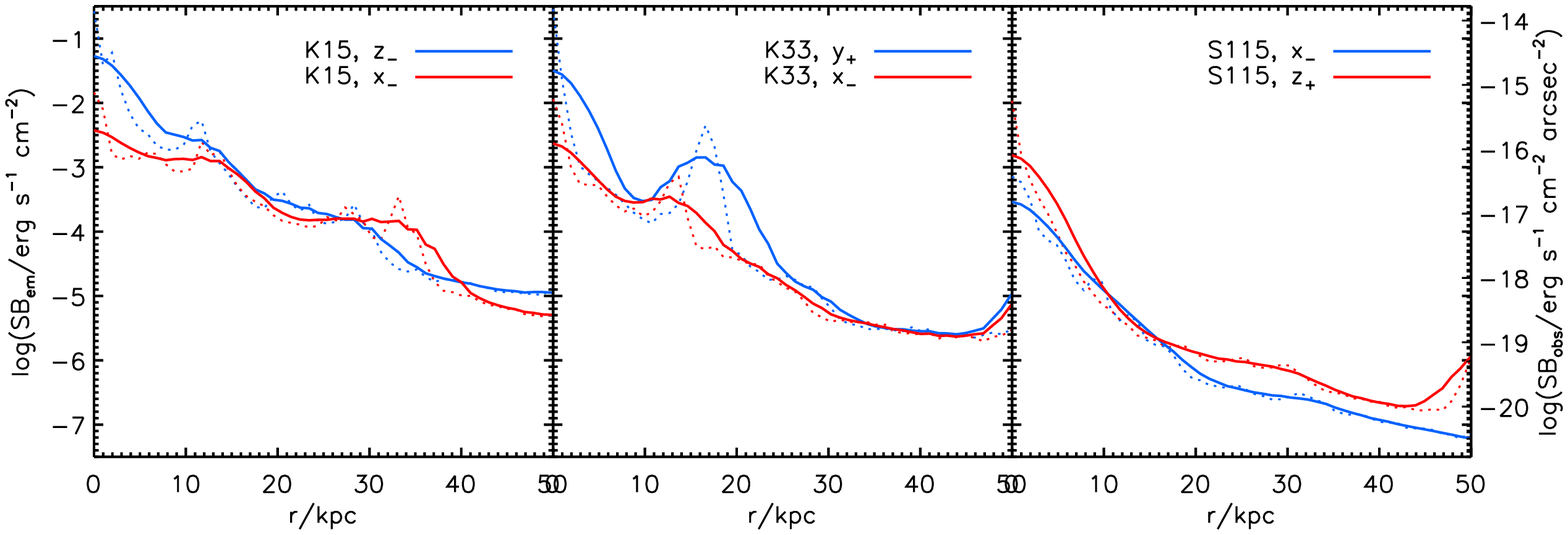}
\caption{Surface brightness (SB) profiles for the inner 50 kpc of the three
         galaxies K15
         (\emph{right}), K33 (\emph{middle}), and S115 (\emph{left}). For each
         galaxy, the SB is showed as observed from two different angles
         (\emph{blue} and \emph{red}, respectively). Both
         the ``true'' SB profiles
         (\emph{dotted})
         and the profiles of the SB convolved with a seeing of $0\farcs8$
         (\emph{solid})
         are shown. Left $y$-axis gives the SB that would be seen by an
         observer at the location of the galaxies, while right $y$-axis gives
         the SB as observed from Earth.}
\label{fig:SBprofiles}
\end{figure*}

Averaging the SB in the azimuthal direction, the SB profiles of the three
galaxies, each as viewed from two different directions, are shown in
Fig.~\ref{fig:SBprofiles}, while Tab.~\ref{tab:SBmax} summarizes the observed
maximum surface brightnesses, SB$_{\mathrm{max}}$.
\begin{deluxetable}{lccc}
\tablecolumns{4}
\tablewidth{0pc}
\tablecaption{Maximum Observed Surface Brightnesses from Different Directions}
\tablehead{
\colhead{Galaxy} & \colhead{K15} & \colhead{K33} & \colhead{S115} \\
}
\startdata                                                                                                
log\,SB$_{\mathrm{max,}x_{+}}$ &         $8.6\times10^{-3}$  &         $6.2\times10^{-3}$  &         $1.3\times10^{-3}$  \\
log\,SB$_{\mathrm{max,}x_{-}}$ & $\mathbf{3.7\times10^{-3}}$ & $\mathbf{2.3\times10^{-3}}$ & $\mathbf{1.5\times10^{-3}}$ \\
log\,SB$_{\mathrm{max,}y_{+}}$ &         $9.9\times10^{-3}$  & $\mathbf{3.2\times10^{-2}}$ &         $7.6\times10^{-4}$  \\
log\,SB$_{\mathrm{max,}y_{-}}$ &         $1.2\times10^{-2}$  &         $2.1\times10^{-2}$  &         $1.2\times10^{-3}$  \\
log\,SB$_{\mathrm{max,}z_{+}}$ &         $4.9\times10^{-2}$  &         $3.4\times10^{-3}$  & $\mathbf{2.9\times10^{-4}}$ \\
log\,SB$_{\mathrm{max,}z_{-}}$ & $\mathbf{5.3\times10^{-2}}$ &         $4.7\times10^{-3}$  &         $4.0\times10^{-4}$  \\
\cline{1-4}\\
Ratio                          &   14.3       &    13.6      &      5.3     \\
\enddata
\tablecomments{Surface brightnesses (SBs, calculated from the images convolved
               with a seeing of $0\farcs8$) are measured in erg s$^{-1}$ cm$^{-2}$,
               at the location of the galaxies.
               The maximally and the minimally observed SB$_{\mathrm{max}}$'s
               for a given galaxy are written in boldface, and the ratios
               between these are given in the lower row.}
\label{tab:SBmax}
\end{deluxetable}
In fact, regarding K33,
SB$_{\mathrm{max,}x_{-}}$ turns out to be smaller than both 
SB$_{\mathrm{max,}z_{-}}$ and SB$_{\mathrm{max,}z_{+}}$, due to the presence in
the line of sight of hydrogen clouds with little star formation causing a
shadowing effect. As seen in Tab.~\ref{tab:SBmax}, the observed
SB of a given galaxy varies with viewing angle by approximately an
order of magnitude.

This result is intriguing in relation to the classification of galaxies.
Galaxies are commonly annotated according to the method by which they are
selected, and one
of the mysteries in the context of galaxy formation and evolution is the
connection between the different types.

As already mentioned, sufficiently high column densities of neutral hydrogen
in the line of sight toward a bright
background source give rise to broad absorption lines in their spectra, and
may be detected as the so-called DLAs.
On the other hand, galaxies with high enough star formation rates (SFRs) may be
detected in
narrowband searches by an excess of their narrowband to continuum flux as
LAEs.

Due to the massive amount of neutral hydrogen,
DLAs are self-shielded against ionizing radiation and may hence be able to
cool sufficiently to initiate star formation.
The assumption that DLAs be progenitors of present-day galaxies therefore
seems reasonable, and is indeed generally accepted. Pursuing this idea, one may
thus image an evolutionary sequence, for instance
DLA $\rightarrow$ LAE $\rightarrow$ LBG \citep[see, e.g.,][]{gaw06a,gaw07}.
However, high-redshift galaxy classification may also be a simple
consequence of a selection effect,
i.e. reflecting the means by which they are probed.
It seems safe to say that the exact relation between the different types
remains unclear.

All three galaxies of the present study contain enough
neutral hydrogen to make them detectable as DLAs in the
spectra of (hypothetical) quasars \citep[see][]{ell07}.
More interestingly, the present results show that while their high SFRs
(K15: 16 $M_\odot$ yr$^{-1}$; K33: 13 $M_\odot$ yr$^{-1}$; S115:
1/2 $M_\odot$ yr$^{-1}$) may make at least the two larger galaxies
detectable as LAEs when viewed from a given direction, it may not be
possible to see them in Ly$\alpha$ from another direction; instead,
it may be possible to observe them as LBGs. This effect demonstrates how
galaxies selected by different means may be connected to each other,
although this obviously has to be quantified through detailed
modeling, including the effect of dust.

Overlaps in the properties of LAEs and LBGs have also been inferred
observationally; \citet{gaw06b} found that more than $80$\% of a sample
of emission-line-selected LAEs have the right $UVB$-colors to be selected as
LBGs. The primary difference between the two populations is the
selection criteria, as only $\sim 10$\% are also brighter than the
$R_{\mathrm{AB}} < 25.5$ ``spectroscopic'' LBG magnitude cut.
Also, when correcting for dust, \citet{gro07} found comparable SFRs for the
two populations.

For high-redshift LAEs, SFRs are inferred almost exclusively from Ly$\alpha$
flux measurements, assuming isotropic luminosity. However, as is evident
from the above discussion, in general the complex morphology of a galaxy
may very well cause a preferred direction of photon escape. For
the three galaxies of the present study, the angular variation
in flux can be as high as a factor of 3.4, 6.2, and 3.3, for K15,
K33, and S115, respectively (here the flux is calculated by integrating
the SB maps over a region of radius $r = 25$ kpc, centered at $r = 0$).
Although not as pronounced as in the case of SB$_{\mathrm{max}}$,
this introduces a considerable source of uncertainty, which may propagate
into estimates of SFRs or into calculations
of the content of dust residing in galaxies.




\subsubsection{Testing Resolution and Interpolation Scheme}
\label{sec:moretests}

In order to check the impact of the resolution of the cosmological simulation
on the results of the Ly$\alpha$ RT, the above RT calculations were also
carried out on the output of cosmological simulations performed at eight times
lower resolution. Furthermore, the procedure by which the
SPH particles were interpolated onto the grid was tested by using the 10
nearest neighboring particles instead of the usual 50 particles, and
running similar RT calculations on the output grids.
Although in both cases the results changed somewhat, there
seems to be no general trend. 
Due to the slightly different evolution of the lo-res galaxies, the precise
configuration of stars and gas clouds will not necessarily be the same, and
thus luminous peaks in the SB maps cannot be expected to coincide exactly.
However, the maximum SBs appear to agree to within a few tens of percents, as
do the slopes and the overall amplitudes of the SB profiles.
The outcome of three such simulations can be
\begin{figure*}
\epsscale{1.2}
\plotone{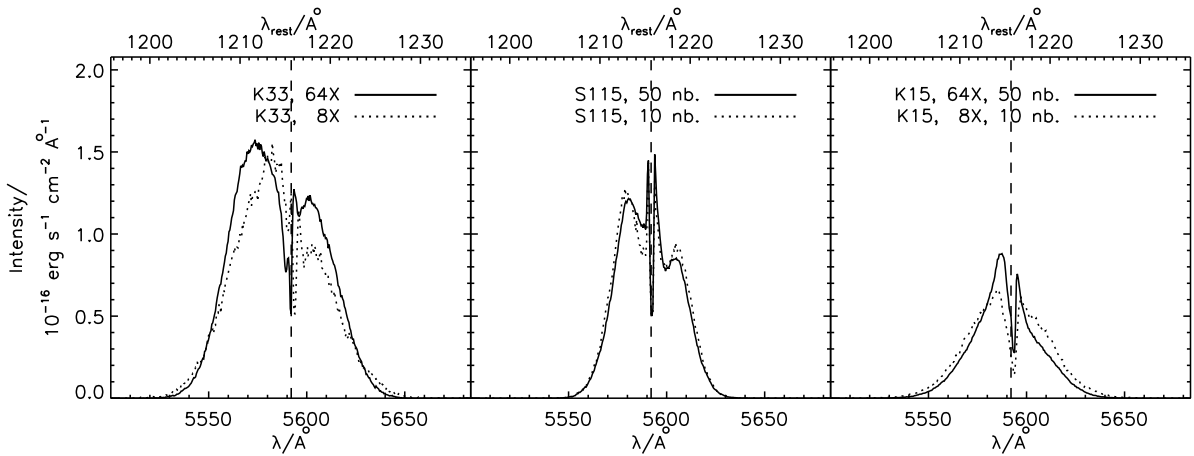}
\caption{Comparison spectra for the resolution test and the interpolation test.
         Left panel shows the spectrum escaping in the negative $y$-direction
         of K33, simulated at high (solid curve) and intermediate (dotted)
         resolution.
         Middle panel shows the spectrum escaping in the negative
         $x$-direction of S115, simulated at ultrahigh (512$\times$)
         resolution, but
         interpolating the physical parameters onto the AMR grid using the
         50 nearest neighboring particles (solid) and the 10 nearest neighbors
         (dotted). Right panel shows the spectrum escaping in the
         negative $z$-direction of K15, simulated at high resolution and
         interpolating from 50 neighbors (solid), compared with intermediate
         resolution/10 neighbors (dotted). The differences
         do not change the results qualitatively. Please note that in the
         middle (right) panel, the intensity has been multiplied (divided) by
         a factor of 10 in order to use the same scale as for all three
         galaxies.}
\label{fig:SpectraComp}
\end{figure*}
\begin{figure*}
\epsscale{1.2}
\plotone{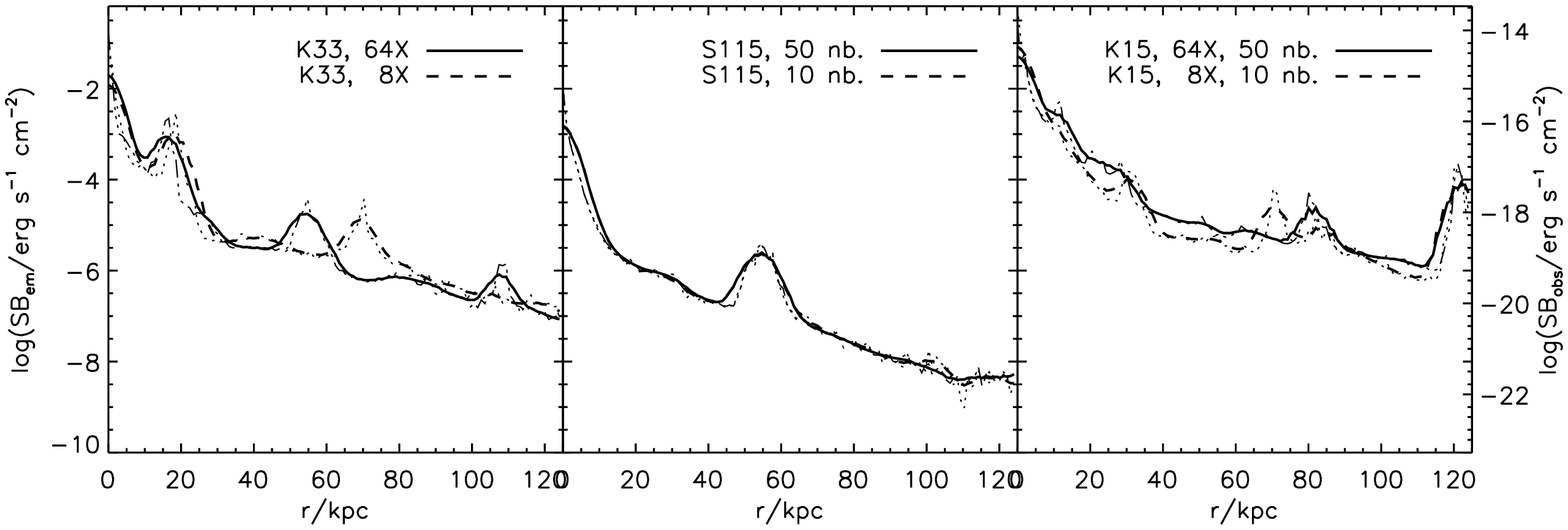}
\caption{Comparison surface brightness (SB) profiles of the six models from
         Fig.~\ref{fig:SpectraComp} at $0\farcs8$ (solid and dashed curves)
         overplotted on the true SB profiles (dotted).
         While performing the simulations at
         different resolutions may shift some of the luminous regions somewhat
         spatially, the maximum SB and the overall slope remain virtually
         unaltered. Moreover, modifying the number of neighboring particles
         used for the interpolation scheme seems unimportant.}
\label{fig:SBprofComp}
\end{figure*}
seen in Figures \ref{fig:SpectraComp} and \ref{fig:SBprofComp}.
%

\section{Summary and Discussion}
\label{sec:disc}

A Monte Carlo Ly$\alpha$ radiative transfer code has been presented, and
subsequently tested against various analytical solutions.
The code is capable of
propagating Ly$\alpha$ radiation on an adaptively refined mesh, with
an arbitrary
distribution of Ly$\alpha$ source emission, gas temperature, density, and
velocity field, and is thus suitable for making predictions about the diffusion
of Ly$\alpha$ radiation in simulated galaxies of arbitrarily high resolution.

A similar non-AMR version of the code has already verified the extendedness
of Ly$\alpha$ surface brightness of $z \simeq 3$--4 compared to the continuum as
being due to the resonant scattering of Ly$\alpha$ \citep{lau07}. In the
present work,
we address the question of the impact of the viewing angle on the observed
properties of galaxies. We find that the anisotropic escape of photons may
cause the maximum observed SB to vary quite a lot --- in the three studied
cases on the average by approximately an order of magnitude ---
while the total observed Ly$\alpha$ flux varies somewhat less; a factor of
3--6. We propose that
this effect may sometimes cause confusion in the classification of galaxies,
in the sense
that the same galaxy could be detected as an LBG from one direction
and a LAE from another. In addition, this angular variation
may act as a source of error when inferring star formation rates.\vspace{2mm}

As yet, the developed code includes no effect of dust.
Considerable amounts of dust have been inferred to reside in LBGs
\citep[e.g.,][]{saw98,cal01,tak04,rig06}. Correcting for dust
increases the typical inferred value of the SFR by a factor of
$\sim5$ \citep{red04}. One may naively expect the
presence of dust to amplify the difference between the SB observed from
different angles of inclination. Nevertheless, taking Ly$\alpha$
transfer into account, it is not obvious how
observations will be affected by dust. In fact, the
Ly$\alpha$-to-continuum ratio may even be boosted if most of the dust resides
in clump together with the neutral gas, since Ly$\alpha$ will scatter
off the clouds while continuum photons get absorbed \citep{neu91,han06}.
Also, since the presence of dust induces the formation of hydrogen molecules,
it may actually lower $n_{\textrm{{\scriptsize H}{\tiny \hspace{.1mm}I}}}$
somewhat, making it easier for the photons to escape.

The transfer of the Ly$\alpha$ photons through the intergalactic medium has not
been modeled either. IGM absorption may remove the blue peak of the
spectrum, since the cosmological expansion will eventually shift it into
resonance of neutral hydrogen in the line of sight, thus approximately halving
the observed flux. The damping wing of the absorption profile may sometimes
extend into the red wing, attenuating the line even more. On the other hand,
the effect of IGM may be substantially reduced, since the line-of-sight
density of H{\sc i} absorbers decrease with decreasing redshift, and by
$z = 3.6$, the Universe is largely ionized.
The implementation of dust and IGM RT will presented in a forthcoming paper.

\acknowledgments
We are very thankful to Toru Yamada for discussing with us the yet unpublished
results of a survey of the SSA22 high-density regions with Subaru FOCAS.
We thank Anja C. Andersen for proofreading and commenting.
The simulations were performed on the facilities provided by the
Danish Center for Scientific Computing.
The Dark Cosmology Centre is funded by the Danish National Research Foundation.
\ \\
\ \\
\ \\
\ \\
\ \\
\ \\
\ \\
\ \\
\ \\


\begin{thebibliography}{}
%
\bibitem[Adams(1972)]{ada72} Adams, T.~F. 1972, \apj, 174, 439
\bibitem[Ahn et al.(2000)]{ahn00} Ahn, S.-H., Lee, H.-W., \& Lee, H.~M. 2000,
    JKAS, 33, 29
\bibitem[Ahn et al.(2001)]{ahn01} Ahn, S.-H., Lee, H.-W., \& Lee, H.~M. 2001,
    \apj, 554, 604
\bibitem[Ahn et al.(2002)]{ahn02} Ahn, S.-H., Lee, H.-W., \& Lee, H.~M. 2002,
    \apj, 567, 922
\bibitem[Ambarzumian(1932)]{amb32} Ambarzumian, V.~A. 1932, \mnras, 93, 50
\bibitem[Auer(1965)]{aue65} Auer, L.~H. 1965, \apj, 153, 783
\bibitem[Avery \& House(1968)]{ave68} Avery, L.~W.~\& House, L.~L. 1968, \apj,
    152, 493
\bibitem[Bonilha et al.(1979)]{bon79} Bonilha, J.~R.~M., Ferch, R., Salpeter,
    E.~E., Slater, G., \& Noerdlinger, P.~D. 1979, \apj, 233, 649
 \bibitem[Bouwens et al.(2004)]{bou04} Bouwens, R. J. et al. 2004, \apj, 616, 79
\bibitem[Calzetti(2001)]{cal01} Calzetti, D. 2001, \pasp, 113, 1449
\bibitem[Cantalupo et al.(2005)]{can05} Cantalupo, S., Porciani, C., Lilly,
    S.~J., \& Miniati, F. 2005, \apj, 628, 61
\bibitem[Caroff et al.(1972)]{car72} Caroff, L.~J., Noerdlinger, P.~D., \&
    Scargle, J.~D. 1972, \apj, 176, 439
\bibitem[Chandrasekhar(1935)]{cha35} Chandrasekhar, S. 1935, \zap, 9, 267
\bibitem[Dijkstra et al.(2006)]{dij06} Dijkstra, M., Haiman, Z., \& Spaans, M.
    2006, \apj, 649, 14
\bibitem[Djorgovski \& Thompson(1992)]{djo92} Djorgovski, S.~\& Thompson, D. J.
    1992, IAUS, 149, 337
\bibitem[Dunlop et al.(2007)]{dun07} Dunlop, J.~S., Le Fevre, O., Franx, M., \&
     Fynbo, J.~P.~U 2007, Principal Investigators of the Ultra-VISTA project
\bibitem[Ellison et al.(2007)]{ell07} Ellison, S. L., Hennawi, J. F.,
    Martin, C. L., \& Sommer-Larsen, J. 2007 \mnras, 378, 801
\bibitem[Fardal et al.(2001)]{far01} Fardal, M.~A., Katz, N., Gardner, J. P.,
    Hernquist, L., Weinberg, D. H., \& Dav\'e, R. 2001, \apj, 562, 605
\bibitem[Feautrier(1964)]{fea64} Feautrier, P. 1964, CR, 258, 3189
\bibitem[Ferland et al.(1998)]{fer98} Ferland, G. J. et al. 1998, PASP, 110, 761
\bibitem[Field(1959)]{fie59} Field, G. 1959, \apj, 129, 551
 \bibitem[Gardner(2006)]{gar06} Gardner, J.~P. 2006, Astronomical Facilities of
     the Next Decade, 26th meeting of the IAU
\bibitem[Gawiser(2006a)]{gaw06a} Gawiser, E. 2006, ASP Conf.~Ser., 352, 177
\bibitem[Gawiser et al.(2006b)]{gaw06b} Gawiser, E. et al.  2006, \apjl 642, L13
\bibitem[Gawiser et al.(2007)]{gaw07} Gawiser, E. et al. 2007, \apj, 671, 278
\bibitem[Gronwall et al.(2007)]{gro07} Gronwall, C. et al. 2007, \apj, 667, 79
\bibitem[Haardt \& Madau(1996)]{haa96} Haardt, F. \& Madau, P. 1996, \apj, 461,
    20
\bibitem[Hamilton(1940)]{ham40} Hamilton, D.~R. 1940, Phys.~Rev, 58, 122
\bibitem[Hansen \& Oh(2006)]{han06} Hansen, M. \& Oh, S.~P. 2006, \mnras, 367,
    979
\bibitem[Harrington(1973)]{har73} Harrington, J.~P. 1973, \mnras, 162, 43
\bibitem[Hartmann et al.(1988)]{har88} Hartmann, L.~W., Huchra, J.~P., Geller,
    M.~J., O'Brien, P., \& Wilson, R. 1988, \apj, 326, 101
\bibitem[Henyey(1940)]{hen40} Henyey, L.~G. 1940, PNAS, 26, 50
 \bibitem[Hu et al.(1998)]{hu98} Hu, E.M., Cowie, L.~L., \& McMahon, R.~G. 1998,
     \apj, 502, 99
 \bibitem[Hu et al.(2004)]{hu04} Hu, E.M., Cowie, L.~L., Capak, P., McMahon,
     R.~G., Hayashino, T., \& Komiyama, Y. 2004, \aj, 127, 563
\bibitem[Hummer(1962)]{hum62} Hummer, D.~G. 1962, \mnras, 125, 21
\bibitem[Iye et al.(2007)]{iye07} Iye, M., Ota, K., \& Kashikawa, N. 2006,
    \baas, 38, 1079
\bibitem[Jefferies \& White(1960)]{jef60} Jefferies, J. T. \& White, O. R.
    1960, \apj, 132, 767
\bibitem[Koelbloed(1956)]{koe56} Koelbloed, D. 1956, BAN, 12, 341
\bibitem[Kollmeier(2006)]{kol06} Kollmeier, J. 2006, Ph.D.~thesis:
    \emph{The Intergalactic Medium: Absorption, Emission, Disruption}
\bibitem[Kroupa(1998)]{kro98} Kroupa, P. 1998, \mnras, 298, 231
\bibitem[Laursen \& Sommer-Larsen(2007)]{lau07} Laursen, P. \& Sommer-Larsen,
    J. 2007, \apjl, 657, 69
\bibitem[Leitherer et al.(1999)]{lei99} Leitherer, C. et al. 1999, \apj, 123, 3
\bibitem[Lia et al.(2002)]{lia02} Lia, C., Portinari, L., \& Carraro, G. 2002,
    \mnras, 330, 821
\bibitem[Loeb \& Rybicki(1999)]{loe99} Loeb, R. \& Rybicki, G.~B. 1999, \apj,
    524, 527
\bibitem[Meier \& Terlevich(1981)]{mei81} Meier, D.~L. \& Terlevich, R. 1981,
    \apj, 246, 109
\bibitem[Miller \& Scalo(1979)]{mil79} Miller, G.~E., Scalo, J.~M. 1979, \apj,
    41, 513
\bibitem[Natta \& Beckwith(1986)]{nat86} Natta, A. \& Beckwith, S. 1986, \aap,
    158, 310
\bibitem[Neufeld(1990)]{neu90} Neufeld, D. 1990, \apj, 350, 216
\bibitem[Neufeld(1991)]{neu91} Neufeld, D. 1991, \apj, 370, L85
\bibitem[Osterbrock(1962)]{ost62} Osterbrock, D.~E. 1962, \apj, 135, 195
 \bibitem[Ouchi et al.(2003)]{ouc03} Ouchi, M. et al. 2003, \apj, 582, 60
\bibitem[Panangia \& Ranieri(1973)]{pan73} Panangia, N. \& Ranieri, M. 1973,
    \aap, 24, 219
\bibitem[Partridge \& Peebles(1967)]{par67} Partridge, R.~B. \& Peebles,
    P.~J.~E. 1967, \apj, 147, 868
\bibitem[Phillips \& M\'esz\'aros(1986)]{phi86} Phillips, K.~C. \&
   M\'esz\'aros, P. 1986, \apj, 310, 284
\bibitem[Pierleoni et al.(2007)]{pie07}
    Pierleoni, M., Maselli, A., \& Ciardi, B. 2007, (arXiv:0712.1159)
\bibitem[Press et al.(1992)]{pre92} Press, W.~H., Teukolsky, S.~A., Vetterling.
    W.~T., \& Flannery, B.~P. 1992, \emph{Numerical Recipes in FORTRAN --- The
    Art of Scientific Computing}, 2nd ed., New York: Cambridge University Press
\bibitem[Razoumov \& Cardall(2005)]{raz05} Razoumov, A.~O. \& Cardall, C. Y.
    2005, \mnras, 362, 1413
\bibitem[Razoumov \& Sommer-Larsen(2006)]{raz06} Razoumov, A.~O. \&
    Sommer-Larsen, J. 2006, \apj, , 651, 81
\bibitem[Razoumov \& Sommer-Larsen(2007)]{raz07} Razoumov, A.~O. \&
    Sommer-Larsen, J. 2007, \apj, , 668, 674
\bibitem[Reddy \& Steidel(2004)]{red04} Reddy, N. A. \& Steidel, Charles C.
    2004, \apj, 603, 13
\bibitem[Rhoads et al.(2000)]{rho00} Rhoads, J.~E., Malhotra, S., Dey, A.,
    Stern, D., Spinrad, H., \& Jannuzi, B.~T. 2000, \apj, 545, 85
\bibitem[Rigopoulou et al.(2006)]{rig06} Rigopoulou, D. et al 2006, \apj, 648,
    81
\bibitem[Salpeter(1955)]{sal55} Salpeter, E. 1955, \apj, 121, 161
\bibitem[Sawicki \& Thompson(2005)]{saw05} Sawicki, M. \& Thompson, D. 2005,
    \apj 635, 100 
\bibitem[Sawicki \& Yee(1998)]{saw98}Sawicki, M \& Yee, H.~K.~C 1998, \aj, 115,
    1329
\bibitem[Sommer-Larsen(2006)]{som06} Sommer-Larsen, J. 2006, \apj, 644, L1
\bibitem[Sommer-Larsen et al.(2003)]{som03} Sommer-Larsen, J., G\"otz, M., \&
    Portinari, L. 2003, \apj, 596, 47
\bibitem[Spitzer(1944)]{spi44} Spitzer, L. 1944, \apj, 99, 1
\bibitem[Steidel et al.(2003)]{ste03} Steidel, C.~C., Adelberger, K.~L.,
    Shapley, A.~E., Pettini, M., Dickinson, M., \& Giavalisco, M. 2003, \apj,
    592, 728S
\bibitem[Stenflo(1976)]{ste76} Stenflo, J.~O. 1976, \aap, 46, 61
\bibitem[Stenflo(1980)]{ste80} Stenflo, J.~O. 1980, \aap, 84, 68
\bibitem[Takeuchi \& Ishii(2004)]{tak04} Takeuchi, T.~T. \& Ishii, T.~T. 2004,
    \aap, 426, 425
\bibitem[Taniguchi et al.(2005)]{tan05} Taniguchi, Y. et al. 2005, \pasj, 57,
    165
\bibitem[Tapken et al.(2004)]{tap04} Tapken, C., Appenzeller, I., Mehlert, D.,
    Noll, S., \& Richling, S. 2004, \aap, 416, 1L    
\bibitem[Tapken et al.(2007)]{tap07} Tapken, C., Appenzeller, I., Noll, S.,
    Richling, S., Heidt, J., Meinköhn, E., Mehlert, D. 2007, \aap, 467, 63
\bibitem[Tasitsiomi(2006a)]{tas06a} Tasitsiomi, A. 2006a, \apj, 645, 792
\bibitem[Tasitsiomi(2006b)]{tas06b} Tasitsiomi, A. 2006b, \apj, 648, 762
\bibitem[Unno(1952a)]{unn52a} Unno, W. 1952a, \pasj, 3, 158
\bibitem[Unno(1952b)]{unn52b} Unno, W. 1952b, \pasj, 4, 100
\bibitem[Unno(1955)]{unn55} Unno, W. 1955, \pasj, 7, 81
\bibitem[Valls-Gabaud(1993)]{val93} Valls-Gabaud, D. 1993, \apj, 419, 7
\bibitem[Venemans et al.(2005)]{ven05} Venemans, B.et al. 2005, \aap, 431, 793
\bibitem[Verhamme et al.(2006)]{ver06} Verhamme, A., Schaerer, D., \& Maselli,
    A. 2006, \aap, 460, 397
\bibitem[Verhamme et al.(2007)]{ver07} Verhamme, A., Schaerer, D., Atek, H.,
    Tapken, C. 2007, ASPC, 380, 97
\bibitem[Yee \& De Robertis(1991)]{yee91} Yee, H. K. C. \& De Robertis, M. M.
    1991, \apj, 381, 386
\bibitem[Yusuf-Sadeh \& Morris(1984)]{yus84} Yusuf-Sadeh, F. \& Morris, M.
    1984, \apj, 278, 186
\bibitem[Zanstra(1949)]{zan49} Zanstra, H. 1949, BAN, 11, 401
\bibitem[Zanstra(1951)]{zan51} Zanstra, H. 1951, BAN, 11, 359
\bibitem[Zheng \& Miralda-Escud\'e(2002)]{zhe02} Zheng, Z. \&
    Miralda-Escud\'e, J. 2002, \apj, 578, 33
%
\end{thebibliography}
\end{document}